\documentclass[twocolumn]{aastex62}
\usepackage{amsmath}
\usepackage{graphicx}
\usepackage{natbib}

\newcommand{\ha}{\hbox{H$\alpha$}}
\newcommand{\hb}{\hbox{H$\beta$}}
\newcommand{\hg}{\hbox{H$\gamma$}}

\newcommand{\oii}{\hbox{[O\,{\sc ii}]}}
\newcommand{\oiii}{\hbox{[O\,{\sc iii}]}}

\shorttitle{Mass-Metallicity Relation}
\shortauthors{Huang et al.}

\begin{document}

\title{The Mass-Metallicity Relation at $z\sim0.8$: Redshift Evolution and Parameter Dependency}

\author{Chi Huang}
\affil{CAS Key Laboratory for Research in Galaxies and Cosmology, Department of Astronomy, University of Science and Technology of China, Hefei 230026, China}
\affil{School of Astronomy and Space Science, University of Science and Technology of China, Hefei 230026, China \email{huangchi@mail.ustc.edu.cn; xkong@ustc.edu.cn}}
\author[0000-0002-6684-3997]{Hu Zou}
\affil{Key Laboratory of Optical Astronomy, National Astronomical Observatories, Chinese Academy of Sciences, Beijing, 100012, China \email{zouhu@nao.cas.cn}} 
\author[0000-0002-7660-2273]{Xu Kong}
\affil{CAS Key Laboratory for Research in Galaxies and Cosmology, Department of Astronomy, University of Science and Technology of China, Hefei 230026, China}
\affil{School of Astronomy and Space Science, University of Science and Technology of China, Hefei 230026, China}
\author{Johan Comparat}
\affil{Max-Planck-Institut für extraterrestrische Physik, Gießenbachstr. 1, D-85748 Garching, Germany}
\author{Zesen Lin}
\author{Yulong Gao}
\author{Zhixiong Liang}
\affil{CAS Key Laboratory for Research in Galaxies and Cosmology, Department of Astronomy, University of Science and Technology of China, Hefei 230026, China}
\affil{School of Astronomy and Space Science, University of Science and Technology of China, Hefei 230026, China}
\author{Timothee Delubac}
\author{Anand Raichoor}
\author{Jean-Paul Kneib} 
\affil{Institute of Physics, Laboratory of Astrophysics, Ecole Polytechnique Fédérale de Lausanne (EPFL), Observatoire de Sauverny, 1290 Versoix, Switzerland}
\author{Donald P. Schneider}
\affil{Department of Astronomy and Astrophysics, The Pennsylvania State University, University Park, PA 16802}
\affil{Institute for Gravitation and the Cosmos, The Pennsylvania State University, University Park, PA 16802}
\author{Xu Zhou}
\affil{Key Laboratory of Optical Astronomy, National Astronomical Observatories, Chinese Academy of Sciences, Beijing, 100012, China} 
\author{Qirong Yuan}
\affil{Department of Physics, Nanjing Normal University, WenYuan Road 1, Nanjing 210046, People's Republic of China}
\author{Matthew A. Bershady}
\affil{Department of Astronomy, University of Wisconsin, 475 North Charter Street, Madison, WI 53706}
\affil{South African Astronomical Observatory, Cape Town, South Africa}



\begin{abstract}
The spectra of emission-line galaxies (ELGs) from the extended Baryon Oscillation Spectroscopic Survey (eBOSS) of the Sloan Digit Sky Survey (SDSS) are used to study the mass-metallicity relation (MZR) at $z\sim0.8$. The selected sample contains about 180,000 massive star-forming galaxies with $0.6 < z < 1.05$ and $9 < {\rm log}(M_{\star}/M_{\odot}) < 12$. The spectra are stacked in bins of different parameters including redshift, stellar mass, star formation rate (SFR), specific star formation rate (sSFR), half-light radius, mass density, and optical color. The average MZR at $z\sim0.83$ has a downward evolution in the MZR from local to high-redshift universe, which is consistent with previous works. At a specified stellar mass, galaxies with higher SFR/sSFR and larger half-light radius have systematically lower metallicity. This behavior is reversed for galaxies with larger mass density and optical color. Among the above physical parameters, the MZR has the most significant dependency on SFR. Our galaxy sample at $0.6<z<1.05$ approximately follows the fundamental metallicity relation (FMR) in the local universe, although the sample inhomogeneity and incompleteness might have effect on our MZR and FMR.
\end{abstract}

\keywords{Galaxies:Emission-line galaxies --- Galaxy masses: Scaling relations --- Galaxy properties: Galaxy abundances}


\section{Introduction\label{intro}}
The gas-phase metallicity is an important physical parameter to study the evolution of galaxies. It reflects the long-term metal enrichment from star-forming activities, regulated by gas inflow, gas outflow, and stellar winds. Metallicity has tight correlations with stellar mass, luminosity, and rotation velocity \citep{Lequeux1979, Rubin1984, Zaritsky1994}, among which the one with stellar mass is the tightest. Stellar mass traces the total amount of long-term star formation, so it is naturally related to metallicity \citep{Lequeux1979, Tremonti2004, Andrews2013}. The relation between stellar mass and gas-phase metallicity (mass-metallicity relation) was first investigated in the local universe and then expanded to redshifts up to $z\sim3.5$ \citep{Tremonti2004, Savaglio2005, Erb2006, Maiolino2008, Zahid2011}. As redshift increases, the MZR shifts downwards, indicating that more evolved galaxies tend to be more metal-rich \citep{Maiolino2008}.

There have been a number of studies about the influence of other physical properties on MZR, including SFR \citep{Mannucci2010}, morphology \citep{SolAlonso2010}, optical color \citep{Yabe2014}, $D_{n}(4000)$ \citep{Lian2015}, gas mass fraction \citep{Hughes2013}, etc. \citet{Ellison2008} selected more than 40,000 galaxies from  SDSS Data Release 4 to study the systematic effect of sSFR and galaxy size (half-light radius) on MZR and found that galaxies with higher sSFR or larger size have systematically lower gas-phase metallicities by up to 0.2 dex. \citet{Mannucci2010} reported a general relation between stellar mass, gas-phase metallicity, and SFR, designated as fundamental metallicity relation (FMR). By introducing a new quantity $\mu = \rm log(M_{\star}) - 0.32\rm{log(SFR)}$, they defined a projection of the FMR that can reduce the metallicity scatter. This FMR remains valid up to $z\sim2.5$, and the redshift evolution in MZR may be due to the difference of SFR of individual galaxies. Many investigations had confirmed this result in the local universe \citep{Lopez2010, Yates2012, Andrews2013} and at high redshift \citep{Cresci2012, Yabe2014, Salim2015}, while some authors expressed reservations regarding the FMR \citep{Sanchez2013, Steidel2014, Sanders2015}. 

The eBOSS program \citep{Dawson2016} is one of the three main surveys of the fourth generation of the Sloan Digital Sky Survey \citep[SDSS-IV;][]{Blanton2017}, and it is designed to explore the expansion history of the Universe throughout 80\% of cosmic time. Although eBOSS is originally designed for cosmology, it provides an unprecedented large number of spectra of star-forming galaxies (SFGs) at medium redshift, which can be used to statistically obtain many physical properties of galaxies in the eBOSS redshift range. The MZR at medium redshift has previously been investigated with relatively small galaxy samples. In this paper, we use a large number of spectra from eBOSS ELGs and compose high signal-to-noise ratio (S/N) spectra by stacking the spectra of single galaxies in bins of different physical properties. These high S/N composite spectra are used to study the MZR at $z\sim 0.8$ and its dependence on different physical properties, which may assist in understanding the galaxy evolution.

The paper is organized as follows. Section \ref{data} describes our spectroscopic samples and corresponding photometric data. Section \ref{reduction} presents the stacking procedure and flux measurements of emission-lines. Section \ref{sec-pars} describes the measurements of different physical parameters. The MZR relation and parameter dependency are analyzed in Section \ref{result}, and Section \ref{summary} is the summary. Throughout this paper, we adopt a flat $\Lambda$CDM model with $H_{0}=70\ \rm km\ s^{-1}\ Mpc^{-1}$, $\Omega_{\Lambda}=0.7$, and $\Omega_{M}=0.3$ The initial mass function (IMF) of \citet{Chabrier2003} is used in this work.

\section{Spectroscopic and photometric data \label{data}}
\subsection{eBOSS survey and ELG sample\label{data1}}
The eBOSS program is one of the three main surveys of SDSS IV, aiming to explore the expansion history of the universe using different spectroscopic galaxy samples. During July 2014 to March 2019, eBOSS obtained spectra of $\sim$183,000 luminous red galaxies over 4600 $\rm deg^{2}$ at $0.6<z<0.8$, $\sim$185,000 ELGs over 1200 $\rm deg^{2}$ at $0.6<z<1.1$, and $\sim$342,000 quasars over 4600 $\rm deg^{2}$ at $0.8<z<3.5$. These surveys use the Sloan Foundation 2.5m Telescope at Apache Point Observatory \citep{Gunn2006}, and the BOSS spectrograph, which covers the wavelength range of $360\sim1000$ nm with 1000 fibers per 7-deg$^2$ plate at a resolution of $R\sim2000$ \citep{Smee2013}. The ELG target selection is based on $grz$-band photometry \citep{Comparat2016,Raichoor2017} from the Dark Energy Camera Legacy Survey \citep[DECaLS][]{Dey2019}, favoring strong $\oii\lambda3727$ emission in the desired redshift range \citep{Comparat2013,Comparat2015}.  The ELG survey covers an area of $\sim620$ $\rm deg^{2}$ in the Southern Galactic Cap (SGC) and $\sim600$ $\rm deg^{2}$ in the Northern Galactic Cap (NGC) at a target density of 200 $\rm deg^{-2}$ in the NGC and 240 $\rm deg^{-2}$ in the SGC, respectively. 

\begin{figure*}
\begin{center}
\includegraphics[width = \textwidth]{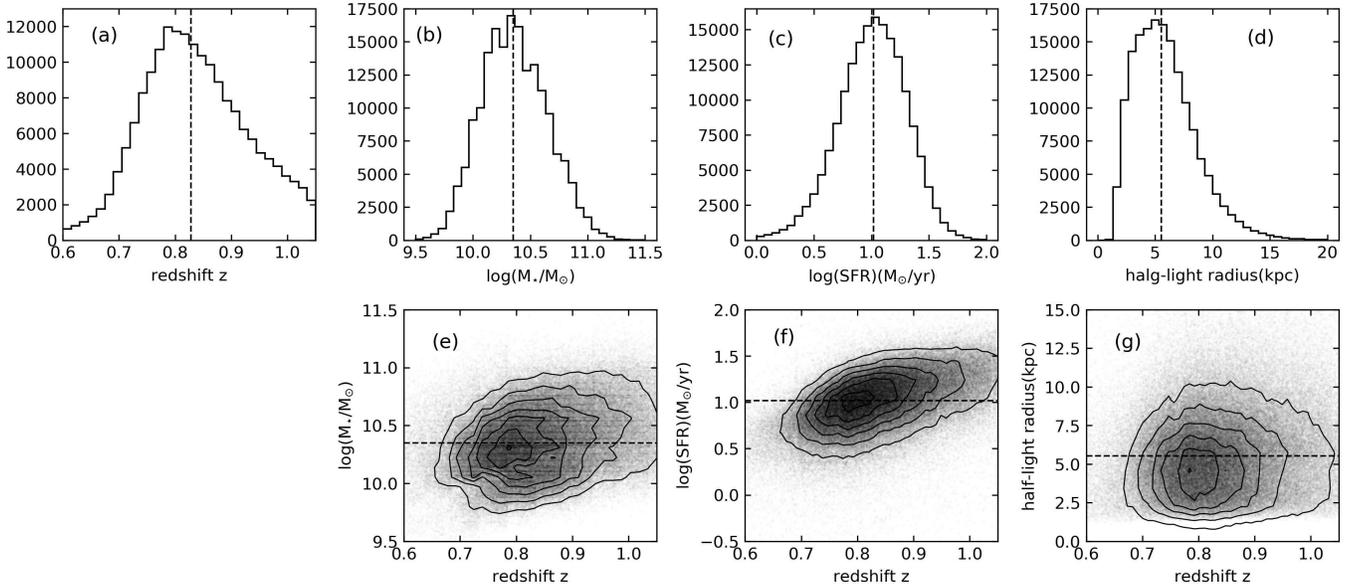}
\caption{(a)--(d): distributions of redshift, stellar mass, SFR, and half-light radius ($R_h$) for our ELG sample. (e)--(g): stellar mass, SFR and $R_h$ as a function of redshift. The solid contours trace the galaxy density distribution. The dashed lines denotes the median values at $z\sim 0.83$, ${\rm log}(M_{\star}/M_{\odot})\sim 10.35$, ${\rm SFR}\sim 10\ M_{\odot}\ \rm \mathrm{yr}^{-1}$, and $R_{h}\sim5.5\ \rm kpc$.}
\label{distribution}
\end{center}
\end{figure*}

The spectra of ELGs are processed by the SDSS spectroscopic pipeline (version of v5\_13\_0), which can reliably derives the redshifts and properties of emission-lines. The ELG observations were completed in 2019 February and the spectra will be released in the SDSS DR16.  A total of 235,123 ELG spectra were obtained. We only select those ELGs with reliably measured redshifts in $0.6 < z < 1.05$, where the redshift quality flags satisfy the conditions of Equation (1) in \cite{Raichoor2017}. The upper limit of $z<1.05$ is set to cover the {\oiii} line. The  stellar mass is within $9<{\rm log}(M_{\star}/M_{\odot})<12$. These constraints produce a sample of 180,020 ELGs for analyses in this paper.  Figure \ref{distribution} presents the properties of our galaxy sample, including the distributions of redshift, stellar mass, SFR, and half-light radius ($R_h$) and the diagrams of mass, SFR, and $R_h$ as a function of redshift. See Section \ref{sec-pars} for the parameter estimations. Our ELG sample has medians of $z\sim0.83$, ${\rm log}(M/M\odot)\sim10.35$, ${\rm SFR}\sim 10\ M_{\odot}\ \mathrm{yr}^{-1}$, and $R_{h}\sim 5.5\ \rm kpc$.

\subsection{DECaLS survey and photometric data\label{data2}}
The DECaLS survey is one of the three recent imaging surveys that are specially designed for the target selections of the Dark Energy Spectroscopic Instrument (DESI) \citep{Dey2019}. DECaLS uses the Dark Energy Camera \citep{Flaugher2015} on the Blanco 4m telescope at the Cerro Tololo Inter-American Observatory to cover an sky area of more than 9,000 deg$^2$ along the equator. It provides $grz$ photometry to fiducial depths of $g = 24.0$, $r=23.4$, and $z=22.5$ mag for $5\sigma$ extended sources. The eBOSS ELG target selection is based on the DECaLS photometry \citep{Raichoor2017}. In addition to the optical imaging, DECaLS integrates historic and latest near-infrared data from the Wide-field Infrared Survey Explorer \citep[WISE][]{Wright2010}. The new coadds of WISE imaging \citep[unWISE;][]{Lang2014} are more than one magnitude deeper than the official ALLWISE. \textit{The Tractor} code \footnote{https://github.com/dstndstn/tractor} \citep{Lang2016} is utilized to calculate the model fluxes of detected sources. Based on the optical and near-infrared fluxes, we obtain a reliable estimation of the stellar mass. 

There are five morphological types: ``PSF" for point sources, ``DEV" for deVaucouleurs profiles, ``EXP" for general exponential profiles, ``REX" for round exponential profiles, and ``COMP" for composite profiles (DEV$+$EXP). The model photometry also provides the shape parameters, among which the half-light radius is taken as an indicator of galaxy size in this paper. The half-light radius is seeing-corrected. 

\section{Spectrum stacking and line measurements\label{reduction}}
\subsection{Stacking procedure\label{reduction1}}
The continuum S/Ns of individual eBOSS spectra in the ELGs are quite low (at $\sim$0.9); in addition, these spectra are significantly contaminated by bright night sky emissions. We stack the spectra using a binning algorithm to obtain high  S/N composite spectra. The stacking procedure is as follows, which is similar to \citet{Zhu2015} and \citet{Lan2018}. 
\begin{enumerate}
\item Each single spectrum is corrected for the Galactic extinction using the extinction law of \citealt{Cardelli1989} and Galactic extinction map of \citet{Schlegel1998}.
\item The spectra are shifted to rest-frame according to measured spectroscopic redshifts.
\item A wavelength grid is constructed, ranging from 2500--5030 $\rm \AA$ with an interval of $0.5\ \rm \AA$. The wavelength range is selected to cover the strong emission-lines bluer than {$\oiii\lambda\lambda5007$}. The step of {0.5 \AA} is selected smaller than the median pixel scale of the eBOSS spectra.  
\item The shifted spectra are linearly interpolated  onto the above wavelength grid. At a given wavelength, the composite spectrum flux is calculated as the median of the individual spectra. Here, the medians instead of the means are adopted in order to reduce the influence of extreme points.   
\item The uncertainties in the pixels in each composite spectrum is estimated from 200 bootstrapping samples.
\end{enumerate}

\begin{figure*}
\begin{center}
\includegraphics[width = 1.0\textwidth]{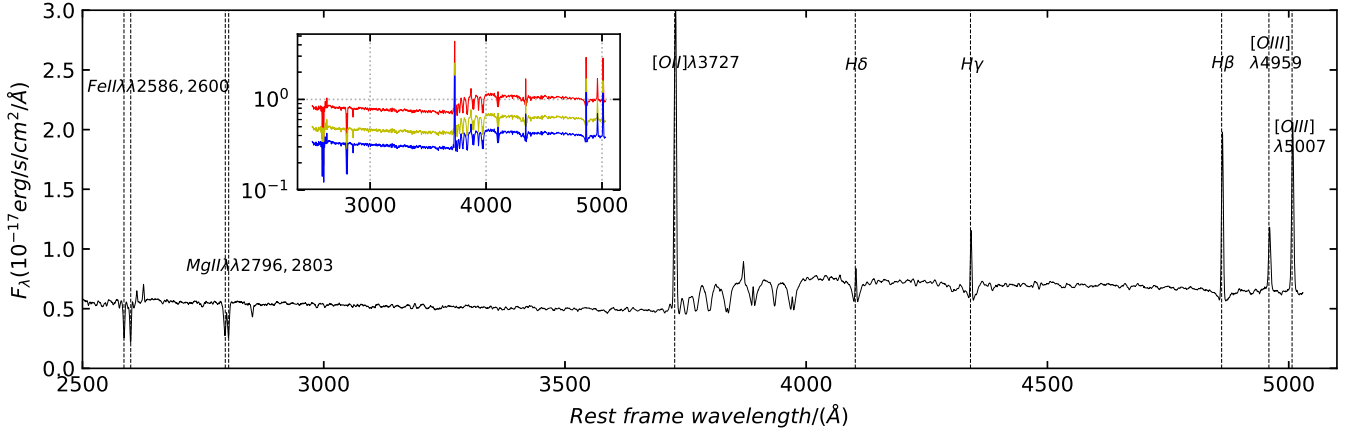}
\caption{The composite spectrum generated with all 180,020 ELGs in $0.6<z<1.05$. The emission and absorption lines are labelled by vertical dashed lines. The inset panel shows the stacked spectra in three narrow redshift bins (red for $0.6< z<0.8$, yellow for $0.8<z<0.9$, and blue for $0.9<z<1.05$). These spectra are shifted vertically by an arbitrary constant value for display purpose.}
\label{spec_total}
\end{center}
\end{figure*}

Figure \ref{spec_total} displays a stacked spectrum composed of all 180,020 ELG spectra at $0.6 < z < 1.05$. This composite spectrum contains several strong emission-lines including $\oii\lambda3727$, $\oiii\lambda4959$, $\oiii\lambda5007$, \hb, \hg, and some absorption features of $\hbox{Mg\,{\sc ii}}\lambda\lambda2796,2803$ and $\hbox{Fe\ {\sc ii}}\lambda\lambda2586,2600$ at the blue end.  

\subsection{Flux measurements of emission-lines\label{reduction2}}
In order to measure the fluxes of emission-lines, we adopt the spectral fitting code of STARLIGHT \citep{CidFernandes2005} to derive the underlying stellar continuum and subtract it from the composite spectrum. The spectral fitting uses 45 single stellar populations from \cite{Bruzual2003} model, \cite{Cardelli1989} extinction law, and \cite{Chabrier2003} IMF. The fluxes and flux errors of emission-lines are calculated by fitting the line profiles with Gaussian functions. Assuming the intrinsic Balmer series ratio of $\hb/\hg=2.137$, we estimate the intrinsic extinction for each composite spectrum. All the fluxes of emission-lines are dereddened using the estimated extinction and \cite{Calzetti2000} attenuation curve.
 
\section{Parameter measurements\label{sec-pars}}
\subsection{Stellar Mass\label{parameter1}}
The availability of deep optical DECaLS $grz$ photometry and near-infrared WISE $W1W2$ photometry permits a reliable estimation of the stellar mass. The eBOSS team has obtained the stellar mass of ELGs through a stellar population synthesis fitting \citep{Raichoor2017}, which fits the spectral energy distribution (SEDs) of ELGs with the FAST code \citep{Kriek2009}. \citet{Raichoor2017} adopts the \cite{Bruzual2003} stellar population models with solar metallicity, \cite{Chabrier2003} IMF, \cite{Kriek2013} dust attenuation law, and exponentially declining star formation history with star formation timescale $\tau$ ranging from 300 Myr to 10 Gyr. During the fit, the redshift is fixed to the spectroscopic redshift. Compared with the COSMOS2015 catalogue \citep{Laigle2016}, which used accurate photometric redshifts and deep optical and near-infrared imaging in 30 bands, \citet{Raichoor2017} showed a good agreement of the stellar mass with a difference of 0.05$\pm$0.21 dex. In this paper, we  adopt the stellar mass derived by the eBOSS team.  A total of 99\% of our selected ELGs having the stellar mass within $8.5<{\rm log}(M_{\star}/M_{\odot})<11.2$. 

\subsection{Metallicity\label{parameter2}}
There are two widely-used methods to measure the gas-phase metallicity of galaxies: one based on electron temperature ($T_{e}$) \citep{Aller1984,Izotov2012,Gao2018} and one based on emission-line ratio of strong lines \citep{McGaugh1991,Zaritsky1994,Pettini2004,Kobulnicky2004,Maiolino2008}. Because \oii$\lambda4363$ becomes too weak to be detected for massive metal-rich galaxies, this line is not detectable in our composite spectra (See Figure \ref{spec_total}). We cannot apply the $T_{e}$ method to estimate the metallicity. Instead, we choose the method described in \citet[][KK04]{Kobulnicky2004}, which is based on four strong emission-lines:
\begin{equation}
R_{23}\equiv\frac{I_\mathrm{[O{\sc II}]\lambda3727}+I_\mathrm{[O{\sc III}]\lambda4959}+I_\mathrm{[O{\sc III}]\lambda5007}}{I_\mathrm{H\beta}}.
\label{R23}
\end{equation} 
However, $R_{23}$ is sensitive to both metallicity and ionization parameter. The relation between $R_{23}$ and metallicity has two branches, so other emission-lines such as $\hbox{[N\,{\sc ii}]}$ and $\hbox{[S\,{\sc ii}]}$ are needed to break the degeneracy \citep{Denicolo2002,Pettini2004}. Unfortunately, these two lines are undetectable in our composite spectra. We adopt the strong line ratio of $O_{32}$ to initially estimate the metallicity:
\begin{equation}
O_{32}\equiv\frac{I_\mathrm{[OIII]\lambda4959}+I_\mathrm{[OIII]\lambda5007}}{I_\mathrm{[OII]\lambda3727}}.
\label{O32}
\end{equation}
This quantity has a monotonic correlation with metallicity but suffers a large dispersion due to its sensitivity to the ionization parameter. \citet[][M08]{Maiolino2008} provided polynomial correlations for the above strong-line metallicity calibrations:
\begin{equation}
\begin{aligned}
{\rm log}R_{23}=0.7462 &- 0.7149x - 0.9401x^{2}\\
& - 0.6154x^{3} - 0.2524x^{4},
\label{logR23}
\end{aligned}
\end{equation}
\begin{equation}
\begin{aligned}
{\rm log}O_{32}=-0.2839 - 1.3881x - 0.3172x^{2}, 
\label{logO32}
\end{aligned}
\end{equation}
where $x=12+{\rm log(O/H)}-8.69$. 
We solve both Equation (\ref{logR23}) and Equation (\ref{logO32}) to find the closest solution of Equation (\ref{logR23}) to the solution of Equation (\ref{logO32}). As a result,  all our composite spectra appear in the upper branch in the metallicity calibration based on $R_{23}$. The relatively high metallicity is reasonable because our selected galaxies are massive. Finally, we apply the polynomial formula for the upper branch specially derived by KK04 (as shown in their Equation (18)) to calculate the metallicity:
\begin{equation}
\begin{aligned}
{\rm 12+log(O/H)}_{\rm upper}\sim&9.11-0.218x-0.0587x^{2}-0.330x^{3} \\
&-0.199x^{4}-y(0.00235\\
&-0.01105x-0.051x^{2}\\
&-0.04085x^{3}-0.003585x^{4}),
\label{KK04}
\end{aligned}
\end{equation}
where $x={\rm log}R_{23}$ and $y={\rm log}O_{32}$. 

\subsection{Star Formation Rate\label{parameter3}}
To explore whether the MZR has a star formation rate (SFR) dependence, we need to estimate the SFR of single eBOSS spectrum. There are two widely-used methods to calculate SFR: one relies on SED fitting and the other is based on the emission-line luminosity. Due to the lack of far-infrared photometry for our sample, the SFR from SED fitting can not be measured reliably, thus we adopt the method based on the the emission-line luminosity.  The luminosity of \ha, $\hb$ and $\oii\lambda3727$ are generally used to infer SFR, among which $\ha$ is the most accurate one. $\ha$ is redshifted to 10,500 $\rm \AA$ at $z\sim0.6$, beyond the maximum wavelength of our spectra. We choose $\hb$ and $\oii\lambda3727$ fluxes as our SFR indicators of single spectrum and adopt the calibrations from \cite{Kennicutt1998} (K98):
\begin{equation}
\begin{aligned}
{\rm SFR}(M_{\odot}\ \rm yr^{-1})&=1.4\times10^{-41}L_\mathrm{[OII]}(\rm erg\ s^{-1})\\
&=7.9\times10^{-42}L_\mathrm{H{\alpha}}(\rm erg\ s^{-1}).
\label{SFR}
\end{aligned}
\end{equation}
The above calibrations are based on the \cite{Salpeter1955} IMF. We use a conversion factor 0.58 as described in \citet{Speagle2014} to convert the SFR to the  \cite{Chabrier2003} IMF, which is adopted throughout this paper. In addition, the intrinsic Balmer line flux ratio of $\ha/\hb=2.86$ is used to convert the $\hb$ luminosity to the $\ha$ one. For dust attenuation, we corrected the fluxes by assuming an average $E(B-V)=0.21$, which is calculated from the overall stack spectra (as shown in Table \ref{tab2}).

\subsection{Half-light Radius\label{parameter4}}
The half-light radius $R_h$ is treated as the galaxy size of our sample. We apply the following restrictions of quality flags in the DECaLS photometric catalogs to obtain reliable photometry:
\begin{equation}
 \begin{aligned}
\mathrm{BRIGHTSTARINBLOB = FALSE}, \\
\mathrm{ALLMASK\_G,R,Z} = 0, \\
\mathrm{FRACMASKED\_G,R,Z} < 0.7, \\
\mathrm{FRACIN\_G,R,Z} < 0.3.
\end{aligned}
\label{equ-quality}
\end{equation}
These quality cuts remove sources with too many masked pixels and avoid photometric contaminations from nearby objects\footnote{\url{http://legacysurvey.org/dr7/files/\#sweep-catalogs}}. In addition, we only choose the ``EXP" and ``REX" types with exponential profiles that are suitable for disk-like galaxies. This approach can avoid possible systematic difference of the $r_h$ measurement due to different galaxy models. 

In order to check the reliability of the DECaLS $R_h$ measurement at $0.6 < z < 1.05$, we cross-match our samples with the public data from Hyper Suprime-Cam Subaru Strategic Program \citep[HSC-SSP][]{Aihara2018}\footnote{https://hsc-release.mtk.nao.ac.jp/doc/}. The imaging of HSC-SSP is almost three magnitudes deeper than the DECaLS and has an exquisite PSF. Comparing the $R_h$ measurements from these two surveys indicates that the two measurements are in good agreement (with a Spearman correlation coefficient of 0.62). 

\section{Results\label{result}}
\subsection{Mass-Metallicity Relation\label{result1}}
The MZR indicates that more massive galaxies tend to be more metal-rich, a trend that holds from local, mid-redshift to distant universe \citep[e.g.,][]{Lequeux1979,Tremonti2004,Savaglio2005,Maiolino2008,Zahid2011,Yabe2014,Lian2015}. We stack the spectra of our sample in six stellar mass bins to obtain six composite spectra with extremely high S/Ns. The median redshift is $z\sim0.83$. We explore the MZR at this redshift.

\begin{figure}
\begin{center}
\includegraphics[width = 0.475\textwidth]{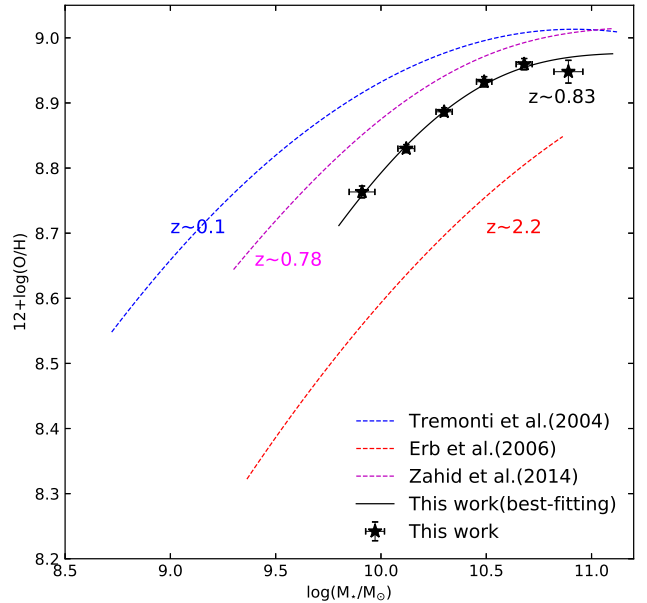}
\caption{The MZRs at four different redshifts. The black solid stars represent our stacked subsamples in six bins of the stellar mass at median redshift of $z = 0.83$. The solid black line is the best-fit curve as formulized in Equation (\ref{curvefit}). The blue dashed curve represents the local MZR at $z\sim0.1$ derived by \citet{Tremonti2004}. The magenta dashed curve presents the MZR at $z\sim0.78$ derived by \citet{Zahid2014}. The red dashed curve shows the MZR at $z\sim2.2$ derived by \citet{Erb2006}.}
\label{MZR}
\end{center}
\end{figure}

As shown in black stars of Figure \ref{MZR}, there is an obvious correlation between mass and metallicity. As the mass increases from $10^{10}M_{\odot}$ to $10^{11}M_{\odot}$, the metallicity 12+log(O/H) rises from 8.77 to 8.96. It appears that there is a change around $10^{10.7} M_{\odot}$, beyond which the metallicity reaches a plateau. We fit this MZR using an analytic form defined by \citet{Zahid2014}:
\begin{equation}
\begin{aligned}
12+{\rm log(O/H)}=Z_{0}+{\rm log}\left[1-{\rm exp}\left(-\left[\frac{M_{\star}}{M_{0}}\right]^{\gamma}\right)\right],
\label{curvefit}
\end{aligned}
\end{equation}
where $Z_{0}$ is the asymptotic metallicity at which the MZR flattens, $M_{0}$ is the turnover mass above which the metallicity asymptotically approches $Z_{0}$, and $\gamma$ is the power-law slope of the MZR for stellar masses $\ll M_{0}$. The best-fit parameters are $Z_{0}=8.977$, $M_{0}=9.961$,  and $\gamma=0.661$ (also see Table \ref{tab1}).

We compare our result with previous studies in Figure \ref{MZR} and show the MZR evolution. There is a remarkable downward evolution trend from $z\sim0.1$ \citep{Tremonti2004}, $z\sim0.78$ \citep{Zahid2014}, $z\sim0.83$ (this work), to $z\sim2.2$ \citep{Erb2006}, which confirms that more evolved galaxies become more metal-rich. All MZRs have been corrected to the same \cite{Chabrier2003} IMF as used in this work. The local MZR at $z\sim0.1$ is from \citet{Tremonti2004}, but it has been transformed by \cite{Savaglio2005} to the same metallicity calibration as ours. The MZR $z\sim2.2$ is derived by \citet{Erb2006} and redetermined in M08, who obtained their own metallicity calibration based on the photoionization model of \cite{Kewley2002}. We apply the calibration of M08 to our sample and find that the average difference of estimated metallicity using the M08 and KK04 calibrations is less than 0.01 dex, suggesting that there is no significant difference in the metallicity estimations between M08 and KK04 as used in this paper. The MZR at  $z\sim0.78$ was derived by \citet{Zahid2014}, who used 50,000 galaxies from the Deep Extragalactic Evolutionary Probe 2 survey \citep[DEEP2][]{Davis2003}. Although \citet{Zahid2014} also adopted the KK04 calibration, they used the calibration formulae as presented in Equation (13) and (17) of KK04. However, we use the calibration formula as described in Equation (18) of KK04, which was considered as the ``best estimate" of the oxygen abundance. There is a systematic offset of -0.082 between the above two calibrations according to Table 1 in KK04. This offset is applied to the metallicity of \citet{Zahid2014} for a consistent comparison. The corresponding MZR curve is indicated in magenta dashed line in Figure \ref{MZR}. The MZR of \citet{Zahid2014} is higher than our measurement, which is reasonable as our median redshift of $z\sim0.83$ is somewhat higher than their value.

\begin{deluxetable*}{ccrccc}
\centering
\tabletypesize{\small}
\tablecaption{Best-fitted parameters of the MZRs in different redshift bins \label{tab1}}
\tablewidth{0pt}
\tablehead{
\colhead{Redshift range} &
\colhead{Median} &
\colhead{$N_\mathrm{gal}$} &
\colhead{$Z_{0}$} &
\colhead{$M_{0}$} &
\colhead{$\gamma$} 
}
\startdata
$0.60\sim1.05$  & 0.83 &  180,020  &  $8.977\pm0.044$  &  $9.961\pm0.018$ & $0.661\pm0.082$\\ 
$0.60\sim0.80$  & 0.75 &  67,960  &  $8.997\pm0.083$  &  $9.929\pm0.031$ & $0.623\pm0.107$\\ 
$0.80\sim0.90$  & 0.84 &  67,292 &  $8.990\pm0.030$  &  $10.054\pm0.012$ & $0.707\pm0.050$\\ 
$0.90\sim1.05$  & 0.95 &   44,768 &  $8.975\pm0.134$  &  $10.072\pm0.050$ & $0.616\pm0.199$
\enddata
\end{deluxetable*}

\begin{figure}[!tb]
\begin{center}
\includegraphics[width = 0.485\textwidth]{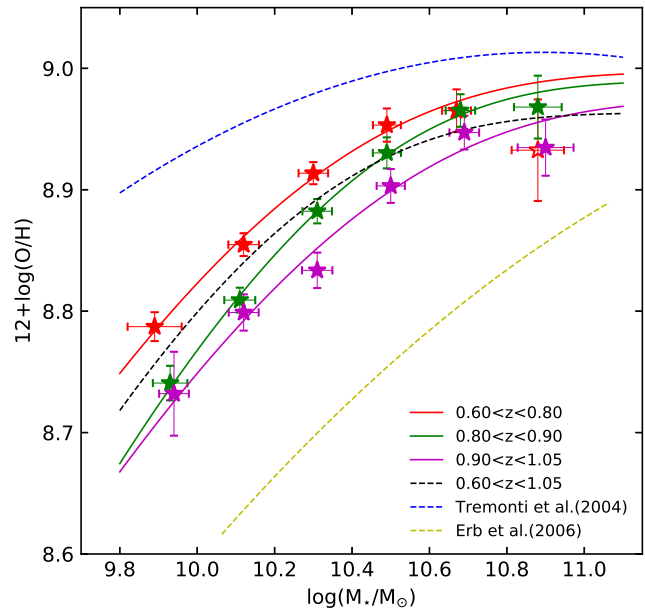}
\caption{MZRs in galaxy samples as a function of redshift. Our galaxy sample is divided into three redshift bins with median redshifts of 0.75, 0.84, and 0.96: $0.60< z <0.80$ (red), $0.80<z<0.90$ (green), and $0.90<z<1.05$ (purple). The MZRs for different redshift bins are shown in solid stars. The corresponding best-fit analytic formulae are indicated with solid lines. The open star marks the most massive point in the lowest redshift bin. It is excluded in the MZR curve fitting due to the low S/N of this bin's composite spectrum. The black dashed line is the overall MZR derived with the total galaxy sample in $0.60<z<1.05$, the same as the black solid line in Figure \ref{MZR}. The blue and yellow dashed lines are the local MZR \citep[]{Tremonti2004} and the $z\sim2.2$ MZR \citep{Erb2006}, which are the same as in Figure \ref{MZR}.}
\label{redshiftevolution}
\end{center}
\end{figure}

We further divide our galaxy sample into different redshift bins to investigate the MZR evolution in our redshift range. Three redshift bins with median redshifts of 0.75, 0.84 and 0.95 are selected: $0.60<z<0.80$, $0.80<z<0.90$, and $0.90<z<1.05$. The MZRs in these redshift ranges are shown in Figure \ref{redshiftevolution}. The best-fit curves with Equation (\ref{curvefit}) are displayed and the corresponding fitted parameters are listed in Table \ref{tab1}. Related physical properties for different stacked spectra are listed in Table \ref{tab2}.  For the lowest redshift bin ($0.60<z<0.80$), the point for the most massive subsample is indicated in open star in Figure \ref{redshiftevolution} as it has unusually low metallicity. This subsample has a smaller number of spectra than other subsamples, and the S/Ns of emission-lines in the composite spectrum are relatively low, resulting in a larger uncertainty in the metallicity estimation. Figure \ref{redshiftevolution} reveals shifted MZRs in three redshift bins. The MZR moves downward as the redshift increases, and at the same time the saturation metallicity $Z_{0}$ becomes lower. Comparing the MZR for the total sample in $0.6 < z < 1.05$ (median $z\sim 0.83$) with the one for the subsample at $0.80<z<0.90$ (median $z \sim 0.84$), the overall MZR suffers more redshift inhomogeneity of the galaxy samples so that the narrower redshift ranges generate more accurate MZRs. 
 
\subsection{Physical Parameter Dependency\label{phydep}}
\begin{figure*}
\begin{center}
\includegraphics[width = 1.0\textwidth]{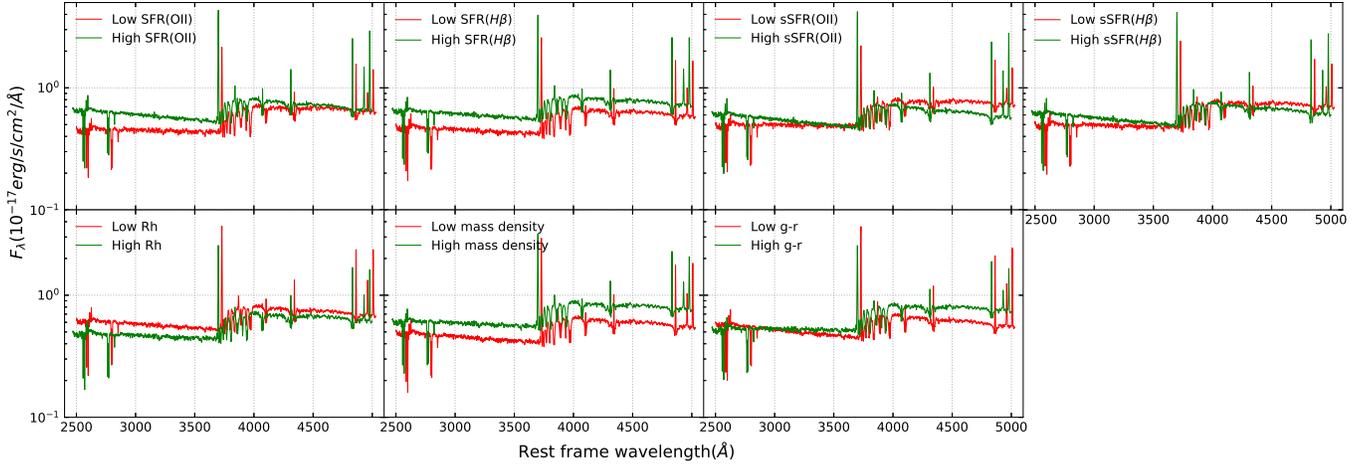}
\caption{Composite spectra stacked in bins of SFR/sSFR(\oii/\hb), half light radius($R_{h}$), mass density, and optical color. The total number of the galaxies in samples are listed in Table \ref{tab2}. The galaxies are selected in a narrow redshift range of $0.8<z<0.9$. The spectra in green are blueshifted by 30 $\rm \AA$ for better visual comparison.}
\label{spec_bins}
\end{center}
\end{figure*}

Our galaxy sample is divided into different bins of SFR (sSFR), half-light radius, mass density, and optical color to investigate their effect on the MZR. The redshift is restricted to $0.8<z<0.9$ to improve the sample redshift homogeneity. Figure \ref{spec_bins} presents the stacked spectra in different parameter bins and Table \ref{tab2} lists the corresponding spectral properties. Figure \ref{spec_bins} has two unsurprising features: (1) the continuum and UV slope of the high-SFR (sSFR) spectrum is higher than the low SFR(sSFR) one, which reflects more active star-forming activity;  (2) the continuum of the high mass density group is lower than the low mass density one. 

\begin{figure}[!tb]
\begin{center}
\includegraphics[width = 0.5\textwidth]{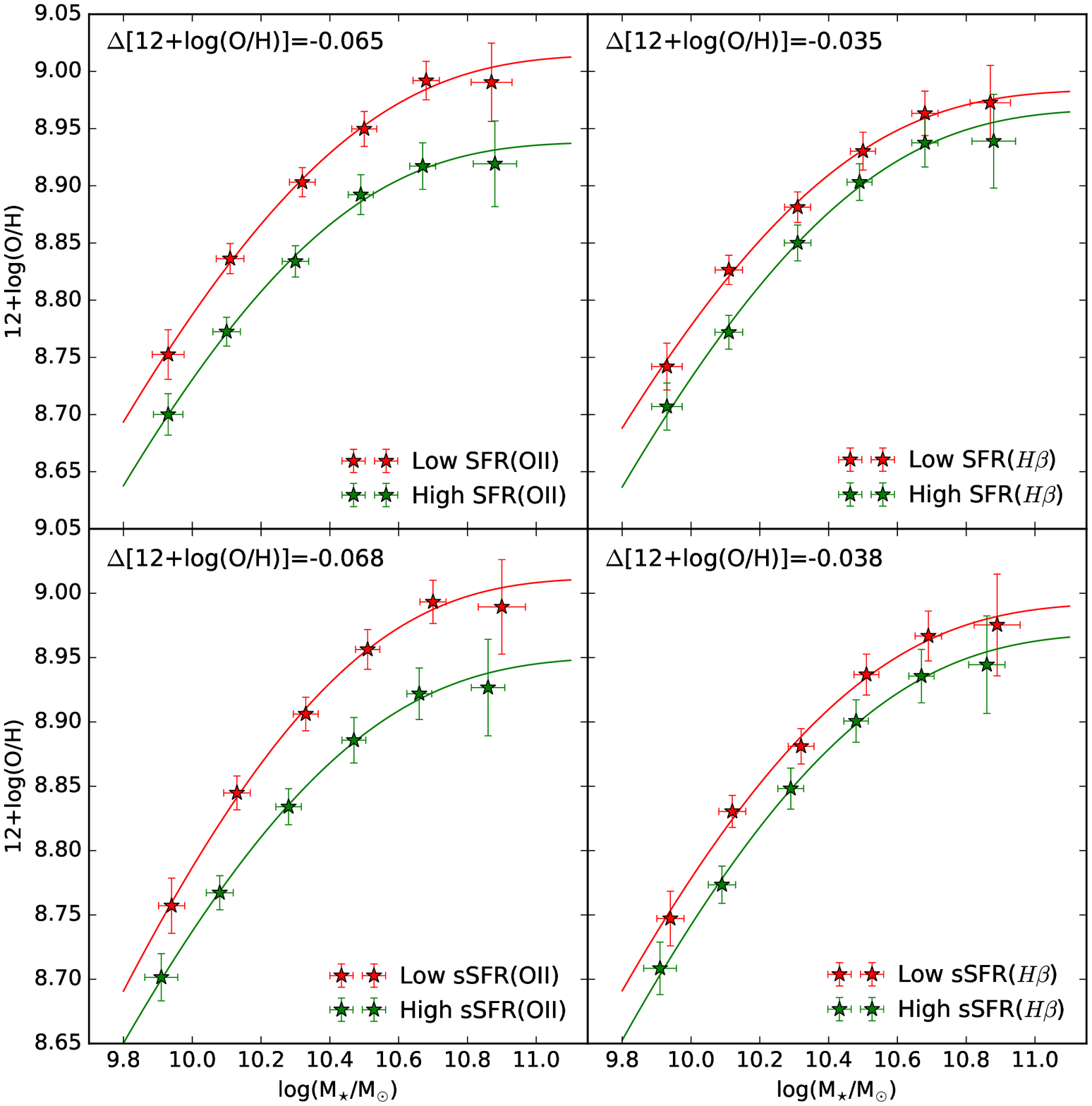}
\caption{MZRs for different bins of SFR and sSFR estimates by either {\oii} or {\hb} luminosity for the $0.8<z<0.9$ sample. Different colors represent different parameter bins. The adopted parameter ranges are presented in Table \ref{tab2}. The average metallicity difference for different parameter bins is annotated in the top-left corner of each panel.}
\label{phyevo1}
\end{center}
\end{figure}

\textbf{Star Formation Rate.} As illustrated in Figure \ref{phyevo1}, the MZR has an obvious deviation between low- and high-SFR (sSFR) bins. The $\oii$ luminosity not only depends on SFR but also correlates with metallicity. The $\hb$ luminosity is a better tracer of the SFR than $\oii$. SFR has been considered as a third parameter of MZR in many previous works \citep[e.g.,][]{Ellison2008,Mannucci2010,Andrews2013}. SFR represents the star-forming activity, implying the information of gas inflow and outflow in galaxies, which are the main physical processes to change the gas-phase metallicity. \citet{Mannucci2010} derived a FMR using local SDSS galaxy sample. The FMR is considered to be a general relation between the mass, metallicity and SFR. \citet{Mannucci2010} introduced a quantity $\mu_{\alpha}=\rm log(M_{\star})-\alpha\rm{log(SFR)}$ to minimize the scatter of MZR ($\alpha=0.32$), and reported that galaxies up to $z\sim2.5$ still follow this relation.

\begin{figure}[!tb]
\begin{center}
\includegraphics[width = 0.47\textwidth]{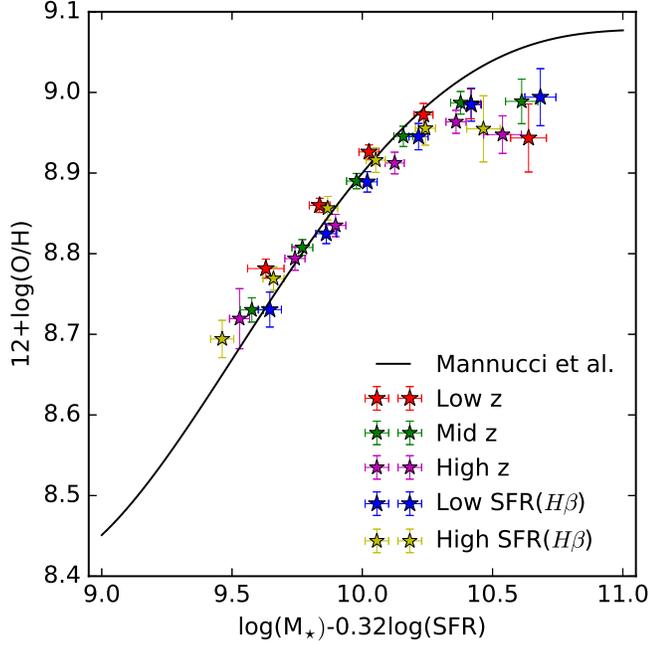}
\caption{FMRs in different redshift and SFR bins for the $0.8<z<0.9$ sample. The solid black line is the local FMR derived by \cite{Mannucci2010}. }
\label{FMR}
\end{center}
\end{figure}

Although the SFRs estimated by $\oii$ and $\hb$ luminosities from single spectrum are not corrected by dust attenuation, the availability of high-S/N lines for $\hb$ and $\hg$ in the stacked spectra enables a calculation of the dust attenuation using the flux ratio of $\hb/\hg$, allowing us to reliably explore the FMR in different SFR bins at $0.8 < z < 0.9$ as well as in different redshift bins.  Figure \ref{FMR} presents our FMRs in the $12+\rm{log(O/H)}$ vs. $\mu$ plane.  We redetermined the metallicity using M08 calibration, which is used in the FMR derived by \cite{Mannucci2010}. The stellar mass and SFR are corrected to \citet{Chabrier2003} IMF. All our FMRs agree with the local FMR derived by \cite{Mannucci2010} within the measurement uncertainties, implying that the FMR at $0.6 < z < 1.05$ still follows the local relation; our work provides a good supplement for FMR at $z\sim0.8$. At $\mu > 10.2$, the metallicity become flatter, which is likely caused by the sample incompleteness.

\begin{figure}[!tb]
\begin{center}
\includegraphics[width = 0.5\textwidth]{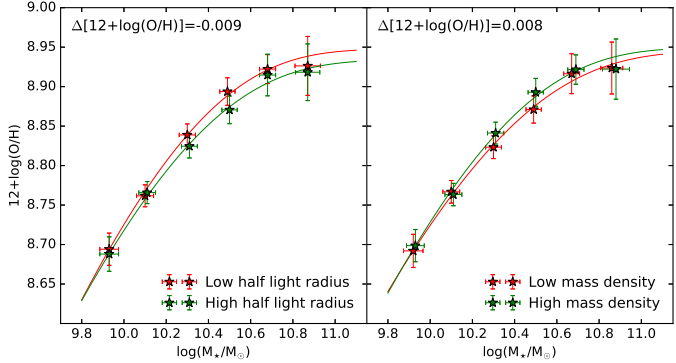}
\caption{MZRs for different bins of $R_h$ and mass density for the $0.8<z<0.9$ sample. Different colors represent different parameter bins. The adopted parameter ranges are presented in Table \ref{tab2}. The average metallicity difference for different parameter bins is annotated in the top-left corner of each panel.}
\label{phyevo2}
\end{center}
\end{figure}

 \textbf{Galaxy size and mass density.}  \cite{Ellison2008} found that the MZR has a dependency on half-light radius in local universe. At a specified stellar mass, the metallicity decreases as the half-light radius increases. Our MZRs for different $R_h$ bins (shown in Figure \ref{phyevo2}) reveal a similar result to \cite{Ellison2008}. We also separate our sample in bins of mass density, which defined as $\Sigma=M_{\star}/\pi R_{h}^{2}$. Figure  \ref{phyevo2} suggests the slight trend that galaxies with higher mass density at a fixed mass have higher metallicity. Higher mass density yields stronger gravitational potential, so that galaxies with high mass density have more ability to keep their metal material from escaping. The correlation between metallicity and mass density may have a more direct physical association than the one between metallicity and half-light radius.
 
\begin{figure}[!tb]
\begin{center}
\includegraphics[width = 0.5\textwidth]{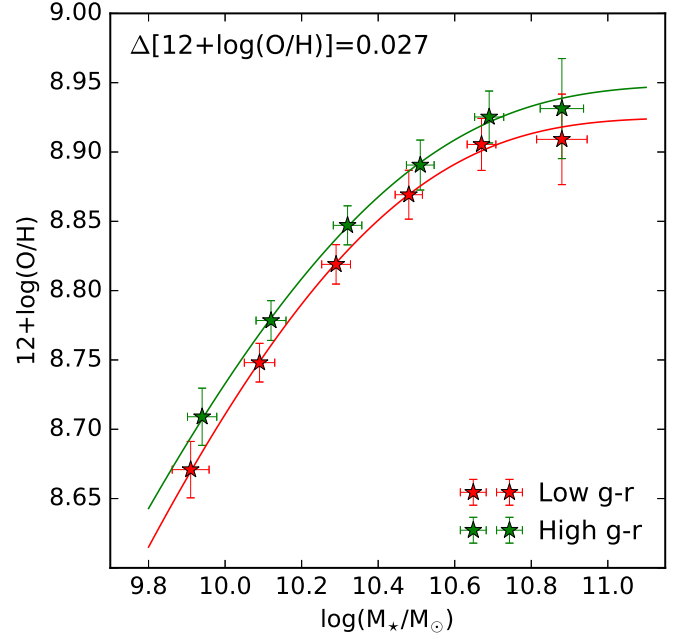}
\caption{MZRs for different bins of $g-r$ color for the $0.8<z<0.9$ sample. Different colors represent different parameter bins. The adopted parameter ranges are presented in Table \ref{tab2}. The average metallicity difference for different color bins is annotated in the top-left corner of each panel.}
\label{phyevo3}
\end{center}
\end{figure}

\textbf{Optical color.}  The optical color to a considerable extent reflects the stellar age, i.e., a redder color means an older age. The MZRs for different color bins in observed $g - r$ are shown in Figure \ref{phyevo3}. The galaxies with redder colors tend to be more metal-rich and hence higher MZRs. The quantity of $D_{n}(4000)$, related to the 4000 $\rm \AA$ break, is defined as the ratio of average fluxes in $\rm4000\sim4100\ \AA$ and $\rm3850\sim3950\ \AA$ \citep{Balogh1999}, and is regarded as a good indicator of galaxy age . \cite{Lian2015} discovered that the MZR has a dependency on $D_{n}(4000)$ at $z\sim1.4$ using a sample of Lyman-break analogues.  The metallicity increases as $D_{n}(4000)$ increases for a specified stellar mass. As shown in Table \ref{tab2}, $D_{n}(4000)$ is larger for redder optical color. Given that galaxies with redder colors in general have larger stellar age and hence more evolved, it is logical that redder galaxies have relatively higher metallicity.

In conclusion, at a specified stellar mass, galaxies with higher SFR/sSFR and half-light radius have lower metallicity, causing the MZR to move downwards. Conversely, galaxies with higher mass density and larger ages have higher metallicity, producing an upward shift in the MZR shift upwards. Our results reflect a possible galaxy evolution picture: galaxies with lower $R_{h}$ at a fixed stellar mass lead to the higher surface mass density triggering higher star formation efficiency (SFE); a long-term higher SFE yields higher present-day metallicity and lower present-day SFR due to the gas depletion. Among the above physical parameters, according to the average $\Delta \rm{[12+log(O/H)]}$ for different parameter bins (shown in left-upper corner in Figure \ref{phyevo1} and \ref{phyevo2}), the MZR has the strongest dependency on the SFR. The significance of the MZR dependency gradually degrades for optical color of $g-r$, half-light radius and mass density.    

\startlongtable
\begin{deluxetable*}{ccccccccccc}
\centering
\tabletypesize{\tiny}
\tablecaption{Physical Properties of Stacked Spectra \label{tab2}}
\tablewidth{0pt}
\tablehead{
 \colhead{redshift range} &\colhead{mass range} & \colhead{redshift} &  \colhead{log($M_{\star}/M_\sun$)} & \colhead{12 + log(O/H)}& \colhead{log(SFR)}& \colhead{$R_{h}$} & \colhead{E(B-V)}& \colhead{$D_{n}(4000)$}& \colhead{$N_\mathrm{gal}$} & \colhead{SNR}\\
 (1) & (2) & (3) & (4) & (5) & (6) & (7) & (8) & (9) & (10) & (11)}
\startdata
\multicolumn{11}{c}{Stacks of overall redshift bin}  \\ 
\tableline
(0.60,1.05) & (9.0,12.0) &$0.83  \pm   0.06  $ &     $  10.35  \pm   0.19  $ &     $   8.89  \pm   0.00  $ &     $   0.98  \pm   0.00  $ &     $   5.52  \pm   1.84  $ &     $   0.21  \pm   0.00  $ &     $   1.18  \pm   0.00  $ &     180020   &   398  \\ 
.. & (9.0,10.0) &$   0.79  \pm   0.05  $ &     $   9.91  \pm   0.06  $ &     $   8.76  \pm   0.01  $ &     $   0.96  \pm   0.00  $ &     $   4.39  \pm   1.62  $ &     $   0.16  \pm   0.00  $ &     $   1.12  \pm   0.00  $ &     16989   &   104  \\ 
.. & (10.0,10.2) &$   0.81  \pm   0.06  $ &     $  10.12  \pm   0.04  $ &     $   8.83  \pm   0.01  $ &     $   0.98  \pm   0.00  $ &     $   4.78  \pm   1.63  $ &     $   0.18  \pm   0.00  $ &     $   1.14  \pm   0.00  $ &     38581   &   173  \\ 
.. & (10.2,10.4) &$   0.82  \pm   0.06  $ &     $  10.30  \pm   0.04  $ &     $   8.89  \pm   0.01  $ &     $   0.99  \pm   0.00  $ &     $   5.45  \pm   1.69  $ &     $   0.21  \pm   0.00  $ &     $   1.17  \pm   0.00  $ &     47983   &   209  \\ 
.. & (10.4,10.6) &$   0.84  \pm   0.06  $ &     $  10.49  \pm   0.04  $ &     $   8.93  \pm   0.01  $ &     $   0.97  \pm   0.01  $ &     $   6.04  \pm   1.82  $ &     $   0.21  \pm   0.01  $ &     $   1.19  \pm   0.00  $ &     37887   &   198  \\ 
.. & (10.6,10.8) &$   0.86  \pm   0.06  $ &     $  10.68  \pm   0.04  $ &     $   8.96  \pm   0.01  $ &     $   0.92  \pm   0.01  $ &     $   6.42  \pm   1.96  $ &     $   0.21  \pm   0.01  $ &     $   1.22  \pm   0.00  $ &     25150   &   169  \\ 
.. & (10.8,12.0) &$   0.90  \pm   0.06  $ &     $  10.89  \pm   0.07  $ &     $   8.95  \pm   0.02  $ &     $   0.94  \pm   0.01  $ &     $   6.74  \pm   2.22  $ &     $   0.22  \pm   0.01  $ &     $   1.25  \pm   0.00  $ &     13430   &   129  \\ 
\tableline
\multicolumn{11}{c}{Stacks of low-redshift bin}\\
\tableline
(0.60,0.80) &(9.0,12.0) &$   0.75  \pm   0.03  $ &     $  10.28  \pm   0.18  $ &     $   8.89  \pm   0.01  $ &     $   0.86  \pm   0.00  $ &     $   5.26  \pm   1.67  $ &     $   0.21  \pm   0.00  $ &     $   1.19  \pm   0.00  $ &     67960   &   249  \\ 
.. & (9.0,10.0) &$   0.75  \pm   0.03  $ &     $   9.89  \pm   0.07  $ &     $   8.79  \pm   0.01  $ &     $   0.81  \pm   0.01  $ &     $   4.55  \pm   1.59  $ &     $   0.13  \pm   0.01  $ &     $   1.14  \pm   0.00  $ &     8955   &   76  \\ 
.. & (10.0,10.2) &$   0.75  \pm   0.03  $ &     $  10.12  \pm   0.04  $ &     $   8.85  \pm   0.01  $ &     $   0.89  \pm   0.01  $ &     $   4.89  \pm   1.50  $ &     $   0.20  \pm   0.01  $ &     $   1.17  \pm   0.00  $ &     16998   &   117  \\ 
.. &  (10.2,10.4)&$   0.75  \pm   0.03  $ &     $  10.30  \pm   0.04  $ &     $   8.91  \pm   0.01  $ &     $   0.86  \pm   0.01  $ &     $   5.42  \pm   1.60  $ &     $   0.21  \pm   0.01  $ &     $   1.19  \pm   0.00  $ &     20538   &   142  \\ 
.. & (10.4,10.6) &$   0.76  \pm   0.03  $ &     $  10.49  \pm   0.04  $ &     $   8.95  \pm   0.01  $ &     $   0.80  \pm   0.01  $ &     $   5.73  \pm   1.77  $ &     $   0.20  \pm   0.01  $ &     $   1.22  \pm   0.00  $ &     13077   &   124  \\ 
.. & (10.6,10.8) &$   0.76  \pm   0.03  $ &     $  10.67  \pm   0.04  $ &     $   8.97  \pm   0.02  $ &     $   0.79  \pm   0.02  $ &     $   5.81  \pm   1.90  $ &     $   0.23  \pm   0.02  $ &     $   1.24  \pm   0.00  $ &     6210   &   89  \\ 
.. & (10.8,12.0) &$   0.76  \pm   0.03  $ &     $  10.88  \pm   0.07  $ &     $   8.93  \pm   0.04  $ &     $   0.76  \pm   0.04  $ &     $   5.56  \pm   2.03  $ &     $   0.25  \pm   0.04  $ &     $   1.28  \pm   0.00  $ &     2182   &   55  \\ 
\tableline
\multicolumn{11}{c}{Stacks of mid-redshift bin}\\
\tableline
(0.80,0.90) & (9.0,12.0) &$   0.84  \pm   0.02  $ &     $  10.35  \pm   0.19  $ &     $   8.88  \pm   0.01  $ &     $   1.04  \pm   0.00  $ &     $   5.55  \pm   1.83  $ &     $   0.22  \pm   0.00  $ &     $   1.17  \pm   0.00  $ &     67292   &   253  \\ 
.. &  (9.0,10.0)&$   0.84  \pm   0.02  $ &     $   9.93  \pm   0.04  $ &     $   8.74  \pm   0.01  $ &     $   1.11  \pm   0.01  $ &     $   4.22  \pm   1.60  $ &     $   0.19  \pm   0.01  $ &     $   1.10  \pm   0.00  $ &     6561   &   66  \\ 
.. & (10.0,10.2) &$   0.85  \pm   0.02  $ &     $  10.11  \pm   0.04  $ &     $   8.81  \pm   0.01  $ &     $   1.06  \pm   0.01  $ &     $   4.69  \pm   1.64  $ &     $   0.18  \pm   0.01  $ &     $   1.12  \pm   0.00  $ &     14450   &   110  \\ 
.. &  (10.2,10.4) &$   0.84  \pm   0.02  $ &     $  10.31  \pm   0.04  $ &     $   8.88  \pm   0.01  $ &     $   1.04  \pm   0.01  $ &     $   5.51  \pm   1.70  $ &     $   0.21  \pm   0.01  $ &     $   1.16  \pm   0.00  $ &     17096   &   128  \\ 
..& (10.4,10.6) &$   0.85  \pm   0.02  $ &     $  10.49  \pm   0.04  $ &     $   8.93  \pm   0.01  $ &     $   1.04  \pm   0.01  $ &     $   6.14  \pm   1.77  $ &     $   0.24  \pm   0.01  $ &     $   1.19  \pm   0.00  $ &     14346   &   126  \\ 
..&  (10.6,10.8) &$   0.85  \pm   0.02  $ &     $  10.68  \pm   0.04  $ &     $   8.97  \pm   0.01  $ &     $   0.95  \pm   0.01  $ &     $   6.44  \pm   1.90  $ &     $   0.24  \pm   0.01  $ &     $   1.23  \pm   0.00  $ &     10273   &   115  \\ 
.. & (10.8,12.0)&$   0.85  \pm   0.02  $ &     $  10.88  \pm   0.06  $ &     $   8.97  \pm   0.03  $ &     $   0.84  \pm   0.02  $ &     $   6.81  \pm   2.22  $ &     $   0.21  \pm   0.02  $ &     $   1.26  \pm   0.00  $ &     4566   &   81  \\ 
\tableline
 \multicolumn{11}{c}{Stacks of high-redshift bin}\\
 \tableline
 (0.90,1.05) & (9.0,12.0) &$ 0.95  \pm   0.03  $ &     $  10.46  \pm   0.19  $ &     $   8.88  \pm   0.01  $ &     $   1.18  \pm   0.00  $ &     $   5.97  \pm   2.03  $ &     $   0.22  \pm   0.00  $ &     $   1.17  \pm   0.00  $ &     44768   &   194  \\ 
 .. & (9.0,10.0) &$   0.93  \pm   0.03  $ &     $   9.94  \pm   0.04  $ &     $   8.73  \pm   0.03  $ &     $   1.28  \pm   0.02  $ &     $   4.26  \pm   1.83  $ &     $   0.21  \pm   0.02  $ &     $   1.10  \pm   0.00  $ &     1473   &   30  \\ 
 ..& (10.0,10.2) &$   0.95  \pm   0.03  $ &     $  10.12  \pm   0.04  $ &     $   8.80  \pm   0.01  $ &     $   1.18  \pm   0.01  $ &     $   4.63  \pm   1.90  $ &     $   0.16  \pm   0.01  $ &     $   1.11  \pm   0.00  $ &     7133   &   72  \\ 
 .. & (10.2,10.4)   &$   0.95  \pm   0.03  $ &     $  10.31  \pm   0.04  $ &     $   8.83  \pm   0.01  $ &     $   1.29  \pm   0.01  $ &     $   5.39  \pm   1.86  $ &     $   0.24  \pm   0.01  $ &     $   1.14  \pm   0.00  $ &     10349   &   91  \\ 
 .. & (10.4,10.6) &$   0.96  \pm   0.03  $ &     $  10.50  \pm   0.04  $ &     $   8.90  \pm   0.01  $ &     $   1.17  \pm   0.01  $ &     $   6.27  \pm   1.91  $ &     $   0.21  \pm   0.01  $ &     $   1.16  \pm   0.00  $ &     10464   &   94  \\ 
 ..& (10.6,10.8) &$   0.96  \pm   0.03  $ &     $  10.69  \pm   0.04  $ &     $   8.95  \pm   0.01  $ &     $   1.03  \pm   0.01  $ &     $   6.79  \pm   2.01  $ &     $   0.18  \pm   0.01  $ &     $   1.20  \pm   0.00  $ &     8667   &   89  \\ 
 .. & (10.8,12.0) &$   0.97  \pm   0.03  $ &     $  10.90  \pm   0.07  $ &     $   8.93  \pm   0.02  $ &     $   1.13  \pm   0.02  $ &     $   7.08  \pm   2.23  $ &     $   0.26  \pm   0.02  $ &     $   1.23  \pm   0.00  $ &     6682   &   86  \\ 
\tableline
\multicolumn{11}{c}{Stacks of low SFR($\oii$) bin (SFR $< 10\ M_{\odot}\ yr^{-1}$)}\\
\tableline
(0.80,0.90) &(9.0,12.0)&$   0.84  \pm   0.02  $ &     $  10.46  \pm   0.18  $ &     $   8.93  \pm   0.01  $ &     $   0.97  \pm   0.01  $ &     $   6.35  \pm   1.88  $ &     $   0.24  \pm   0.01  $ &     $   1.21  \pm   0.00  $ &     33034   &   168  \\ 
.. & (9.0,10.0)  &$   0.83  \pm   0.02  $ &     $   9.93  \pm   0.05  $ &     $   8.75  \pm   0.02  $ &     $   1.07  \pm   0.01  $ &     $   5.08  \pm   1.78  $ &     $   0.21  \pm   0.01  $ &     $   1.12  \pm   0.00  $ &     3260   &   42  \\ 
.. & (10.0,10.2) &$   0.84  \pm   0.02  $ &     $  10.11  \pm   0.04  $ &     $   8.84  \pm   0.01  $ &     $   0.95  \pm   0.01  $ &     $   5.47  \pm   1.74  $ &     $   0.17  \pm   0.01  $ &     $   1.14  \pm   0.00  $ &     7176   &   70  \\ 
.. & (10.2,10.4)&$   0.83  \pm   0.02  $ &     $  10.32  \pm   0.04  $ &     $   8.90  \pm   0.01  $ &     $   0.99  \pm   0.01  $ &     $   6.11  \pm   1.75  $ &     $   0.23  \pm   0.01  $ &     $   1.17  \pm   0.00  $ &     8418   &   81  \\ 
.. &  (10.4,10.6) &$   0.84  \pm   0.02  $ &     $  10.50  \pm   0.04  $ &     $   8.95  \pm   0.02  $ &     $   1.02  \pm   0.01  $ &     $   6.54  \pm   1.81  $ &     $   0.28  \pm   0.01  $ &     $   1.21  \pm   0.00  $ &     7015   &   81  \\ 
.. & (10.6,10.8) &$   0.84  \pm   0.02  $ &     $  10.68  \pm   0.04  $ &     $   8.99  \pm   0.02  $ &     $   0.91  \pm   0.02  $ &     $   6.84  \pm   1.91  $ &     $   0.25  \pm   0.02  $ &     $   1.25  \pm   0.00  $ &     4973   &   75  \\ 
.. & (10.8,12.0) &$   0.85  \pm   0.02  $ &     $  10.87  \pm   0.06  $ &     $   8.99  \pm   0.03  $ &     $   0.89  \pm   0.03  $ &     $   7.30  \pm   2.23  $ &     $   0.28  \pm   0.04  $ &     $   1.28  \pm   0.00  $ &     2190   &   54  \\ 
\tableline
 \multicolumn{11}{c}{Stacks of high SFR($\oii$) bin (SFR $> 10\ M_{\odot}\ yr^{-1}$)}\\
 \tableline
(0.80,0.90) &(9.0,12.0) &$   0.85  \pm   0.02  $ &     $  10.24  \pm   0.17  $ &     $   8.81  \pm   0.01  $ &     $   1.16  \pm   0.00  $ &     $   4.73  \pm   1.60  $ &     $   0.24  \pm   0.00  $ &     $   1.14  \pm   0.00  $ &     33034   &   191  \\ 
..  & (9.0,10.0)  &$   0.84  \pm   0.02  $ &     $   9.93  \pm   0.04  $ &     $   8.70  \pm   0.02  $ &     $   1.21  \pm   0.01  $ &     $   3.63  \pm   1.19  $ &     $   0.20  \pm   0.01  $ &     $   1.09  \pm   0.00  $ &     3260   &   53  \\ 
..  & (10.0,10.2)&$   0.85  \pm   0.02  $ &     $  10.10  \pm   0.04  $ &     $   8.77  \pm   0.01  $ &     $   1.14  \pm   0.01  $ &     $   4.02  \pm   1.37  $ &     $   0.19  \pm   0.01  $ &     $   1.10  \pm   0.00  $ &     7177   &   90  \\ 
.. & (10.2,10.4) &$   0.85  \pm   0.02  $ &     $  10.30  \pm   0.04  $ &     $   8.83  \pm   0.01  $ &     $   1.18  \pm   0.01  $ &     $   4.87  \pm   1.54  $ &     $   0.24  \pm   0.01  $ &     $   1.13  \pm   0.00  $ &     8419   &   103  \\ 
.. & (10.4,10.6) &$   0.85  \pm   0.02  $ &     $  10.49  \pm   0.04  $ &     $   8.89  \pm   0.02  $ &     $   1.14  \pm   0.01  $ &     $   5.72  \pm   1.68  $ &     $   0.25  \pm   0.01  $ &     $   1.16  \pm   0.00  $ &     7016   &   100  \\ 
.. & (10.6,10.8)&$   0.85  \pm   0.02  $ &     $  10.67  \pm   0.04  $ &     $   8.92  \pm   0.02  $ &     $   1.14  \pm   0.01  $ &     $   5.99  \pm   1.84  $ &     $   0.30  \pm   0.02  $ &     $   1.20  \pm   0.00  $ &     4974   &   90  \\ 
..  &  (10.8,12.0)  &$   0.86  \pm   0.02  $ &     $  10.88  \pm   0.06  $ &     $   8.92  \pm   0.04  $ &     $   0.99  \pm   0.03  $ &     $   6.17  \pm   2.11  $ &     $   0.25  \pm   0.03  $ &     $   1.23  \pm   0.00  $ &     2190   &   61  \\ 
\tableline
 \multicolumn{11}{c}{Stacks of low SFR ($\hb$) bin (SFR $< 5\ M_{\odot}\ yr^{-1}$)}\\
 \tableline
(0.80,0.90) & (9.0,12.0) &$   0.84  \pm   0.02  $ &     $  10.36  \pm   0.19  $ &     $   8.89  \pm   0.01  $ &     $   0.81  \pm   0.00  $ &     $   5.99  \pm   1.85  $ &     $   0.13  \pm   0.00  $ &     $   1.18  \pm   0.00  $ &     29792   &   159  \\ 
..& (9.0,10.0) &$   0.83  \pm   0.02  $ &     $   9.93  \pm   0.04  $ &     $   8.74  \pm   0.02  $ &     $   0.89  \pm   0.01  $ &     $   4.66  \pm   1.70  $ &     $   0.11  \pm   0.01  $ &     $   1.11  \pm   0.00  $ &     2960   &   42  \\ 
.. & (10.0,10.2) &$   0.84  \pm   0.02  $ &     $  10.11  \pm   0.04  $ &     $   8.83  \pm   0.01  $ &     $   0.78  \pm   0.01  $ &     $   5.18  \pm   1.68  $ &     $   0.06  \pm   0.01  $ &     $   1.13  \pm   0.00  $ &     6626   &   69  \\ 
.. & (10.2,10.4) &$   0.84  \pm   0.02  $ &     $  10.31  \pm   0.04  $ &     $   8.88  \pm   0.01  $ &     $   0.91  \pm   0.01  $ &     $   5.91  \pm   1.71  $ &     $   0.17  \pm   0.01  $ &     $   1.16  \pm   0.00  $ &     7662   &   80  \\ 
.. & (10.4,10.6) &$   0.84  \pm   0.02  $ &     $  10.50  \pm   0.04  $ &     $   8.93  \pm   0.02  $ &     $   0.89  \pm   0.01  $ &     $   6.51  \pm   1.77  $ &     $   0.19  \pm   0.01  $ &     $   1.20  \pm   0.00  $ &     6283   &   78  \\ 
.. &  (10.6,10.8)&$   0.85  \pm   0.02  $ &     $  10.68  \pm   0.04  $ &     $   8.96  \pm   0.02  $ &     $   0.82  \pm   0.02  $ &     $   6.78  \pm   1.92  $ &     $   0.19  \pm   0.02  $ &     $   1.24  \pm   0.00  $ &     4356   &   70  \\ 
.. & (10.8,12.0) &$   0.86  \pm   0.02  $ &     $  10.87  \pm   0.06  $ &     $   8.97  \pm   0.03  $ &     $   0.58  \pm   0.03  $ &     $   7.23  \pm   2.25  $ &     $   0.09  \pm   0.03  $ &     $   1.26  \pm   0.00  $ &     1905   &   50  \\ 
 \tableline
  \multicolumn{11}{c}{Stacks of high SFR ($\hb$) bin (SFR $> 5\ M_{\odot}\ yr^{-1}$)}\\
   \tableline
(0.80,0.90)&(9.0,12.0)  &$   0.85  \pm   0.02  $ &     $  10.33  \pm   0.18  $ &     $   8.85  \pm   0.01  $ &     $   1.37  \pm   0.00  $ &     $   4.90  \pm   1.68  $ &     $   0.35  \pm   0.00  $ &     $   1.15  \pm   0.00  $ &     29793   &   190  \\ 
..& (9.0,10.0) &$   0.84  \pm   0.02  $ &     $   9.93  \pm   0.04  $ &     $   8.71  \pm   0.02  $ &     $   1.46  \pm   0.01  $ &     $   3.74  \pm   1.35  $ &     $   0.33  \pm   0.01  $ &     $   1.10  \pm   0.00  $ &     2960   &   50  \\ 
.. &  (10.0,10.2) &$   0.85  \pm   0.02  $ &     $  10.11  \pm   0.04  $ &     $   8.77  \pm   0.01  $ &     $   1.41  \pm   0.01  $ &     $   4.10  \pm   1.44  $ &     $   0.33  \pm   0.01  $ &     $   1.11  \pm   0.00  $ &     6626   &   85  \\ 
.. & (10.2,10.4) &$   0.85  \pm   0.02  $ &     $  10.31  \pm   0.04  $ &     $   8.85  \pm   0.02  $ &     $   1.38  \pm   0.01  $ &     $   4.90  \pm   1.57  $ &     $   0.35  \pm   0.01  $ &     $   1.14  \pm   0.00  $ &     7662   &   98  \\ 
.. &  (10.4,10.6) &$   0.85  \pm   0.02  $ &     $  10.49  \pm   0.04  $ &     $   8.90  \pm   0.02  $ &     $   1.37  \pm   0.01  $ &     $   5.58  \pm   1.69  $ &     $   0.38  \pm   0.01  $ &     $   1.17  \pm   0.00  $ &     6283   &   95  \\ 
.. & (10.6,10.8) &$   0.85  \pm   0.02  $ &     $  10.68  \pm   0.04  $ &     $   8.94  \pm   0.02  $ &     $   1.37  \pm   0.02  $ &     $   5.92  \pm   1.77  $ &     $   0.41  \pm   0.02  $ &     $   1.21  \pm   0.00  $ &     4356   &   84  \\ 
..& (10.8,12.0) &$   0.85  \pm   0.02  $ &     $  10.88  \pm   0.06  $ &     $   8.94  \pm   0.04  $ &     $   1.30  \pm   0.03  $ &     $   6.05  \pm   2.07  $ &     $   0.41  \pm   0.03  $ &     $   1.24  \pm   0.00  $ &     1906   &   59  \\ 
\tableline
 \multicolumn{11}{c}{Stacks of low sSFR ($\oii$) bin (sSFR $< 0.5\ \mathrm{Gyr}^{-1}$)}\\
 \tableline
 (0.80,0.90) & (9.0,12.0) &$   0.84  \pm   0.02  $ &     $  10.55  \pm   0.14  $ &     $   8.94  \pm   0.01  $ &     $   0.99  \pm   0.01  $ &     $   6.40  \pm   1.88  $ &     $   0.25  \pm   0.01  $ &     $   1.21  \pm   0.00  $ &     33034   &   181  \\ 
.. & (9.0,10.0) &$   0.83  \pm   0.02  $ &     $   9.94  \pm   0.04  $ &     $   8.76  \pm   0.02  $ &     $   1.08  \pm   0.01  $ &     $   5.04  \pm   1.76  $ &     $   0.22  \pm   0.01  $ &     $   1.12  \pm   0.00  $ &     3260   &   43  \\ 
.. & (10.0,10.2) &$   0.84  \pm   0.02  $ &     $  10.13  \pm   0.04  $ &     $   8.84  \pm   0.01  $ &     $   0.93  \pm   0.01  $ &     $   5.48  \pm   1.73  $ &     $   0.15  \pm   0.01  $ &     $   1.14  \pm   0.00  $ &     7176   &   71  \\ 
..& (10.2,10.4) &$   0.83  \pm   0.02  $ &     $  10.33  \pm   0.04  $ &     $   8.91  \pm   0.01  $ &     $   0.98  \pm   0.01  $ &     $   6.13  \pm   1.75  $ &     $   0.22  \pm   0.01  $ &     $   1.17  \pm   0.00  $ &     8418   &   82  \\ 
.. & (10.4,10.6) &$   0.84  \pm   0.02  $ &     $  10.51  \pm   0.04  $ &     $   8.96  \pm   0.02  $ &     $   0.98  \pm   0.01  $ &     $   6.55  \pm   1.80  $ &     $   0.26  \pm   0.01  $ &     $   1.21  \pm   0.00  $ &     7015   &   81  \\ 
.. & (10.6,10.8)&$   0.84  \pm   0.02  $ &     $  10.70  \pm   0.04  $ &     $   8.99  \pm   0.02  $ &     $   0.90  \pm   0.02  $ &     $   6.84  \pm   1.92  $ &     $   0.25  \pm   0.02  $ &     $   1.25  \pm   0.00  $ &     4973   &   75  \\ 
.. & (10.8,12.0) &$   0.85  \pm   0.02  $ &     $  10.90  \pm   0.07  $ &     $   8.99  \pm   0.04  $ &     $   0.91  \pm   0.04  $ &     $   7.31  \pm   2.27  $ &     $   0.28  \pm   0.04  $ &     $   1.28  \pm   0.00  $ &     2190   &   53  \\ 
\tableline
\multicolumn{11}{c}{Stacks of high sSFR ($\oii$) bin (sSFR $> 0.5\ \mathrm{Gyr}^{-1}$)}\\
\tableline
(0.80,0.90)& (9.0,12.0) &$   0.85  \pm   0.02  $ &     $  10.16  \pm   0.12  $ &     $   8.81  \pm   0.01  $ &     $   1.14  \pm   0.00  $ &     $   4.70  \pm   1.58  $ &     $   0.22  \pm   0.00  $ &     $   1.13  \pm   0.00  $ &     33034   &   177  \\ 
.. & (9.0,10.0) &$   0.84  \pm   0.02  $ &     $   9.91  \pm   0.05  $ &     $   8.70  \pm   0.02  $ &     $   1.19  \pm   0.01  $ &     $   3.63  \pm   1.24  $ &     $   0.19  \pm   0.01  $ &     $   1.09  \pm   0.00  $ &     3260   &   52  \\ 
..& (10.0,10.2) &$   0.85  \pm   0.02  $ &     $  10.08  \pm   0.04  $ &     $   8.77  \pm   0.01  $ &     $   1.15  \pm   0.01  $ &     $   4.01  \pm   1.37  $ &     $   0.19  \pm   0.01  $ &     $   1.10  \pm   0.00  $ &     7177   &   89  \\ 
.. &(10.2,10.4) &$   0.85  \pm   0.02  $ &     $  10.28  \pm   0.04  $ &     $   8.83  \pm   0.01  $ &     $   1.16  \pm   0.01  $ &     $   4.84  \pm   1.52  $ &     $   0.23  \pm   0.01  $ &     $   1.13  \pm   0.00  $ &     8419   &   102  \\ 
..& (10.4,10.6)&$   0.85  \pm   0.02  $ &     $  10.47  \pm   0.03  $ &     $   8.89  \pm   0.02  $ &     $   1.17  \pm   0.01  $ &     $   5.71  \pm   1.69  $ &     $   0.27  \pm   0.01  $ &     $   1.16  \pm   0.00  $ &     7016   &   100  \\ 
.. & (10.6,10.8) &$   0.85  \pm   0.02  $ &     $  10.66  \pm   0.04  $ &     $   8.92  \pm   0.02  $ &     $   1.10  \pm   0.01  $ &     $   5.99  \pm   1.83  $ &     $   0.28  \pm   0.02  $ &     $   1.20  \pm   0.00  $ &     4974   &   89  \\ 
.. & (10.8,12.0) &$   0.86  \pm   0.02  $ &     $  10.86  \pm   0.05  $ &     $   8.93  \pm   0.04  $ &     $   0.97  \pm   0.03  $ &     $   6.25  \pm   2.06  $ &     $   0.24  \pm   0.03  $ &     $   1.23  \pm   0.00  $ &     2190   &   61  \\ 
 \tableline
\multicolumn{11}{c}{Stacks of low sSFR ($\hb$) bin (sSFR $< 0.2\ \mathrm{Gyr}^{-1}$)}\\
 \tableline
(0.80,0.90)& (9.0,12.0) &$   0.85  \pm   0.02  $ &     $  10.52  \pm   0.16  $ &     $   8.93  \pm   0.01  $ &     $   0.90  \pm   0.01  $ &     $   6.27  \pm   1.85  $ &     $   0.19  \pm   0.01  $ &     $   1.20  \pm   0.00  $ &     29792   &   170  \\ 
.. & (9.0,10.0) &$   0.83  \pm   0.02  $ &     $   9.94  \pm   0.04  $ &     $   8.75  \pm   0.02  $ &     $   0.89  \pm   0.01  $ &     $   4.64  \pm   1.69  $ &     $   0.10  \pm   0.01  $ &     $   1.11  \pm   0.00  $ &     2960   &   42  \\ 
.. & (10.0,10.2)  &$   0.84  \pm   0.02  $ &     $  10.12  \pm   0.04  $ &     $   8.83  \pm   0.01  $ &     $   0.78  \pm   0.01  $ &     $   5.23  \pm   1.68  $ &     $   0.06  \pm   0.01  $ &     $   1.13  \pm   0.00  $ &     6626   &   69  \\ 
.. & (10.2,10.4) &$   0.84  \pm   0.02  $ &     $  10.32  \pm   0.04  $ &     $   8.88  \pm   0.01  $ &     $   0.95  \pm   0.01  $ &     $   5.93  \pm   1.71  $ &     $   0.19  \pm   0.01  $ &     $   1.16  \pm   0.00  $ &     7662   &   80  \\ 
..& (10.4,10.6) &$   0.85  \pm   0.02  $ &     $  10.51  \pm   0.04  $ &     $   8.94  \pm   0.02  $ &     $   0.86  \pm   0.01  $ &     $   6.53  \pm   1.78  $ &     $   0.17  \pm   0.01  $ &     $   1.20  \pm   0.00  $ &     6283   &   78  \\ 
..&  (10.6,10.8)&$   0.85  \pm   0.02  $ &     $  10.69  \pm   0.04  $ &     $   8.97  \pm   0.02  $ &     $   0.83  \pm   0.02  $ &     $   6.80  \pm   1.93  $ &     $   0.19  \pm   0.02  $ &     $   1.24  \pm   0.00  $ &     4356   &   70  \\ 
.. & (10.8,12.0)&$   0.86  \pm   0.02  $ &     $  10.89  \pm   0.07  $ &     $   8.98  \pm   0.04  $ &     $   0.61  \pm   0.03  $ &     $   7.22  \pm   2.26  $ &     $   0.11  \pm   0.04  $ &     $   1.27  \pm   0.00  $ &     1905   &   50  \\ 
 \tableline
 \multicolumn{11}{c}{Stacks of high sSFR ($\hb$) bin (sSFR $> 0.2 \mathrm{Gyr}^{-1}$)}\\
 \tableline
 (0.80,0.90)& (9.0,12.0)  &$   0.85  \pm   0.02  $ &     $  10.18  \pm   0.14  $ &     $   8.82  \pm   0.01  $ &     $   1.28  \pm   0.00  $ &     $   4.65  \pm   1.59  $ &     $   0.29  \pm   0.00  $ &     $   1.14  \pm   0.00  $ &     29793   &   173  \\ 
.. & (9.0,10.0) &$   0.84  \pm   0.02  $ &     $   9.91  \pm   0.05  $ &     $   8.71  \pm   0.02  $ &     $   1.44  \pm   0.01  $ &     $   3.75  \pm   1.37  $ &     $   0.32  \pm   0.01  $ &     $   1.10  \pm   0.00  $ &     2960   &   49  \\ 
.. & (10.0,10.2)  &$   0.85  \pm   0.02  $ &     $  10.09  \pm   0.04  $ &     $   8.77  \pm   0.01  $ &     $   1.37  \pm   0.01  $ &     $   4.08  \pm   1.44  $ &     $   0.31  \pm   0.01  $ &     $   1.11  \pm   0.00  $ &     6626   &   85  \\ 
..& (10.2,10.4) &$   0.85  \pm   0.02  $ &     $  10.29  \pm   0.04  $ &     $   8.85  \pm   0.02  $ &     $   1.37  \pm   0.01  $ &     $   4.87  \pm   1.57  $ &     $   0.34  \pm   0.01  $ &     $   1.14  \pm   0.00  $ &     7662   &   97  \\ 
..& (10.4,10.6) &$   0.85  \pm   0.02  $ &     $  10.48  \pm   0.04  $ &     $   8.90  \pm   0.02  $ &     $   1.38  \pm   0.01  $ &     $   5.56  \pm   1.67  $ &     $   0.38  \pm   0.01  $ &     $   1.17  \pm   0.00  $ &     6283   &   94  \\ 
.. & (10.6,10.8) &$   0.85  \pm   0.02  $ &     $  10.67  \pm   0.04  $ &     $   8.94  \pm   0.02  $ &     $   1.37  \pm   0.02  $ &     $   5.91  \pm   1.77  $ &     $   0.41  \pm   0.02  $ &     $   1.21  \pm   0.00  $ &     4356   &   84  \\ 
.. & (10.8,12.0)&$   0.85  \pm   0.02  $ &     $  10.86  \pm   0.05  $ &     $   8.94  \pm   0.04  $ &     $   1.24  \pm   0.03  $ &     $   6.09  \pm   2.06  $ &     $   0.37  \pm   0.03  $ &     $   1.24  \pm   0.00  $ &     1906   &   58  \\ 
\tableline
\multicolumn{11}{c}{Stacks of low half light radius bin ($R_h < 5.5$ kpc)}\\
\tableline
(0.80,0.90)& (9.0,12.0) &$   0.84  \pm   0.02  $ &     $  10.27  \pm   0.18  $ &     $   8.87  \pm   0.01  $ &     $   1.07  \pm   0.00  $ &     $   3.88  \pm   0.71  $ &     $   0.22  \pm   0.00  $ &     $   1.16  \pm   0.00  $ &     30675   &   180  \\ 
.. &(9.0,10.0) &$   0.84  \pm   0.02  $ &     $   9.93  \pm   0.04  $ &     $   8.74  \pm   0.02  $ &     $   1.14  \pm   0.01  $ &     $   2.98  \pm   0.49  $ &     $   0.19  \pm   0.01  $ &     $   1.10  \pm   0.00  $ &     2920   &   47  \\
.. & (10.0,10.2) &$   0.85  \pm   0.02  $ &     $  10.10  \pm   0.04  $ &     $   8.81  \pm   0.01  $ &     $   1.09  \pm   0.01  $ &     $   3.31  \pm   0.56  $ &     $   0.18  \pm   0.01  $ &     $   1.12  \pm   0.00  $ &     6583   &   82  \\ 
.. &  (10.2,10.4)&$   0.84  \pm   0.02  $ &     $  10.30  \pm   0.04  $ &     $   8.89  \pm   0.01  $ &     $   1.09  \pm   0.01  $ &     $   3.95  \pm   0.69  $ &     $   0.22  \pm   0.01  $ &     $   1.15  \pm   0.00  $ &     7899   &   96  \\ 
.. & (10.4,10.6)&$   0.84  \pm   0.02  $ &     $  10.49  \pm   0.04  $ &     $   8.94  \pm   0.02  $ &     $   1.06  \pm   0.01  $ &     $   4.56  \pm   0.78  $ &     $   0.25  \pm   0.01  $ &     $   1.19  \pm   0.00  $ &     6637   &   95  \\ 
..& (10.6,10.8) &$   0.85  \pm   0.02  $ &     $  10.68  \pm   0.04  $ &     $   8.97  \pm   0.02  $ &     $   1.00  \pm   0.02  $ &     $   4.74  \pm   0.85  $ &     $   0.26  \pm   0.02  $ &     $   1.23  \pm   0.00  $ &     4686   &   85  \\ 
.. & (10.8,12.0) &$   0.85  \pm   0.02  $ &     $  10.87  \pm   0.06  $ &     $   8.98  \pm   0.04  $ &     $   0.72  \pm   0.04  $ &     $   4.68  \pm   0.94  $ &     $   0.14  \pm   0.04  $ &     $   1.25  \pm   0.00  $ &     1948   &   58  \\ 
\tableline
\multicolumn{11}{c}{Stacks of high half light radius bin ($R_h > 5.5$ kpc)}\\
\tableline
(0.80,0.90)&(9.0,12.0) &$   0.85  \pm   0.02  $ &     $  10.43  \pm   0.18  $ &     $   8.90  \pm   0.01  $ &     $   0.99  \pm   0.01  $ &     $   7.47  \pm   1.47  $ &     $   0.21  \pm   0.01  $ &     $   1.19  \pm   0.00  $ &     30676   &   165  \\ 
..&(9.0,10.0) &$   0.83  \pm   0.02  $ &     $   9.93  \pm   0.04  $ &     $   8.74  \pm   0.02  $ &     $   1.06  \pm   0.01  $ &     $   5.91  \pm   1.44  $ &     $   0.18  \pm   0.01  $ &     $   1.11  \pm   0.00  $ &     2921   &   41  \\ 
.. &  (10.0,10.2)&$   0.84  \pm   0.02  $ &     $  10.11  \pm   0.04  $ &     $   8.82  \pm   0.01  $ &     $   0.96  \pm   0.01  $ &     $   6.44  \pm   1.39  $ &     $   0.14  \pm   0.01  $ &     $   1.13  \pm   0.00  $ &     6584   &   68  \\ 
..& (10.2,10.4) &$   0.84  \pm   0.02  $ &     $  10.31  \pm   0.04  $ &     $   8.87  \pm   0.01  $ &     $   1.02  \pm   0.01  $ &     $   7.23  \pm   1.38  $ &     $   0.21  \pm   0.01  $ &     $   1.16  \pm   0.00  $ &     7900   &   80  \\ 
..& (10.4,10.6) &$   0.85  \pm   0.02  $ &     $  10.50  \pm   0.04  $ &     $   8.92  \pm   0.02  $ &     $   1.02  \pm   0.01  $ &     $   7.92  \pm   1.40  $ &     $   0.24  \pm   0.01  $ &     $   1.19  \pm   0.00  $ &     6638   &   79  \\ 
..& (10.6,10.8)&$   0.85  \pm   0.02  $ &     $  10.68  \pm   0.04  $ &     $   8.96  \pm   0.03  $ &     $   0.95  \pm   0.02  $ &     $   8.41  \pm   1.46  $ &     $   0.24  \pm   0.02  $ &     $   1.23  \pm   0.00  $ &     4686   &   73  \\ 
..&  (10.8,12.0)  &$   0.86  \pm   0.02  $ &     $  10.87  \pm   0.06  $ &     $   8.97  \pm   0.04  $ &     $   0.92  \pm   0.03  $ &     $   9.16  \pm   1.68  $ &     $   0.25  \pm   0.03  $ &     $   1.26  \pm   0.00  $ &     1949   &   51  \\ 
\tableline
\multicolumn{11}{c}{Stacks of low mass density bin ($\rm log(\Sigma/(M_{\odot} kpc^{-2})) < 6.5$)}\\ 
\tableline
(0.80,0.90)& (9.0,12.0) &$   0.84  \pm   0.02  $ &     $  10.28  \pm   0.17  $ &     $   8.86  \pm   0.01  $ &     $   1.02  \pm   0.00  $ &     $   7.18  \pm   1.72  $ &     $   0.21  \pm   0.00  $ &     $   1.16  \pm   0.00  $ &     30675   &   159  \\ 
..&(9.0,10.0) &$   0.83  \pm   0.02  $ &     $   9.92  \pm   0.05  $ &     $   8.74  \pm   0.02  $ &     $   1.04  \pm   0.01  $ &     $   5.92  \pm   1.45  $ &     $   0.17  \pm   0.01  $ &     $   1.11  \pm   0.00  $ &     2920   &   41  \\ 
..&  (10.0,10.2)&$   0.84  \pm   0.02  $ &     $  10.10  \pm   0.04  $ &     $   8.82  \pm   0.01  $ &     $   0.96  \pm   0.01  $ &     $   6.44  \pm   1.40  $ &     $   0.14  \pm   0.01  $ &     $   1.13  \pm   0.00  $ &     6583   &   68  \\ 
..&  (10.2,10.4) &$   0.84  \pm   0.02  $ &     $  10.30  \pm   0.04  $ &     $   8.87  \pm   0.01  $ &     $   1.01  \pm   0.01  $ &     $   7.23  \pm   1.40  $ &     $   0.20  \pm   0.01  $ &     $   1.16  \pm   0.00  $ &     7899   &   80  \\ 
..& (10.4,10.6) &$   0.85  \pm   0.02  $ &     $  10.49  \pm   0.04  $ &     $   8.92  \pm   0.02  $ &     $   1.01  \pm   0.01  $ &     $   7.92  \pm   1.41  $ &     $   0.23  \pm   0.01  $ &     $   1.19  \pm   0.00  $ &     6637   &   78  \\ 
..& (10.6,10.8)&$   0.85  \pm   0.02  $ &     $  10.67  \pm   0.04  $ &     $   8.97  \pm   0.03  $ &     $   0.92  \pm   0.02  $ &     $   8.41  \pm   1.48  $ &     $   0.22  \pm   0.02  $ &     $   1.23  \pm   0.00  $ &     4686   &   72  \\ 
..&  (10.8,12.0)  &$   0.86  \pm   0.02  $ &     $  10.86  \pm   0.05  $ &     $   8.97  \pm   0.03  $ &     $   0.90  \pm   0.03  $ &     $   9.15  \pm   1.72  $ &     $   0.24  \pm   0.03  $ &     $   1.26  \pm   0.00  $ &     1948   &   51  \\ 
\tableline
\multicolumn{11}{c}{Stacks of high mass density bin ($\rm log(\Sigma/(M_{\odot} kpc^{-2})) > 6.5$)}\\ 
\tableline
(0.80,0.90)& (9.0,12.0) &$   0.85  \pm   0.02  $ &     $  10.45  \pm   0.19  $ &     $   8.91  \pm   0.01  $ &     $   1.04  \pm   0.01  $ &     $   4.01  \pm   1.13  $ &     $   0.22  \pm   0.01  $ &     $   1.18  \pm   0.00  $ &     30676   &   194  \\ 
.. & (9.0,10.0) &$   0.84  \pm   0.02  $ &     $   9.93  \pm   0.04  $ &     $   8.75  \pm   0.02  $ &     $   1.13  \pm   0.01  $ &     $   2.98  \pm   0.51  $ &     $   0.18  \pm   0.01  $ &     $   1.10  \pm   0.00  $ &     2921   &   47  \\ 
.. & (10.0,10.2) &$   0.85  \pm   0.02  $ &     $  10.11  \pm   0.04  $ &     $   8.81  \pm   0.01  $ &     $   1.09  \pm   0.01  $ &     $   3.31  \pm   0.58  $ &     $   0.18  \pm   0.01  $ &     $   1.12  \pm   0.00  $ &     6584   &   82  \\ 
.. & (10.2,10.4)  &$   0.84  \pm   0.02  $ &     $  10.31  \pm   0.04  $ &     $   8.89  \pm   0.01  $ &     $   1.09  \pm   0.01  $ &     $   3.95  \pm   0.71  $ &     $   0.22  \pm   0.01  $ &     $   1.15  \pm   0.00  $ &     7900   &   96  \\ 
.. & (10.4,10.6) &$   0.84  \pm   0.02  $ &     $  10.50  \pm   0.04  $ &     $   8.94  \pm   0.02  $ &     $   1.08  \pm   0.01  $ &     $   4.56  \pm   0.81  $ &     $   0.26  \pm   0.01  $ &     $   1.19  \pm   0.00  $ &     6638   &   95  \\ 
..& (10.6,10.8)  &$   0.85  \pm   0.02  $ &     $  10.69  \pm   0.04  $ &     $   8.97  \pm   0.02  $ &     $   1.01  \pm   0.02  $ &     $   4.74  \pm   0.87  $ &     $   0.27  \pm   0.02  $ &     $   1.23  \pm   0.00  $ &     4686   &   85  \\ 
.. & (10.8,12.0) &$   0.85  \pm   0.02  $ &     $  10.88  \pm   0.06  $ &     $   8.97  \pm   0.04  $ &     $   0.74  \pm   0.04  $ &     $   4.68  \pm   1.01  $ &     $   0.16  \pm   0.04  $ &     $   1.25  \pm   0.00  $ &     1949   &   59  \\ 
\tableline
 \multicolumn{11}{c}{Stacks of low $g-r$ color bin ($g - r < 0.63$)}\\ 
 \tableline
(0.80,0.90)& (9.0,12.0)&$   0.85  \pm   0.02  $ &     $  10.17  \pm   0.16  $ &     $   8.82  \pm   0.01  $ &     $   1.05  \pm   0.00  $ &     $   5.07  \pm   1.77  $ &     $   0.19  \pm   0.00  $ &     $   1.14  \pm   0.00  $ &     33646   &   170  \\ 
..&(9.0,10.0)&$   0.84  \pm   0.02  $ &     $   9.91  \pm   0.05  $ &     $   8.72  \pm   0.02  $ &     $   1.14  \pm   0.01  $ &     $   4.10  \pm   1.62  $ &     $   0.19  \pm   0.01  $ &     $   1.10  \pm   0.00  $ &     3280   &   47  \\ 
..&  (10.0,10.2)  &$   0.85  \pm   0.02  $ &     $  10.09  \pm   0.04  $ &     $   8.80  \pm   0.01  $ &     $   1.02  \pm   0.01  $ &     $   4.66  \pm   1.66  $ &     $   0.15  \pm   0.01  $ &     $   1.11  \pm   0.00  $ &     7225   &   78  \\ 
..&  (10.2,10.4)  &$   0.85  \pm   0.02  $ &     $  10.29  \pm   0.04  $ &     $   8.87  \pm   0.01  $ &     $   1.03  \pm   0.01  $ &     $   5.56  \pm   1.71  $ &     $   0.19  \pm   0.01  $ &     $   1.15  \pm   0.00  $ &     8548   &   90  \\ 
..& (10.4,10.6)&$   0.86  \pm   0.02  $ &     $  10.48  \pm   0.04  $ &     $   8.92  \pm   0.02  $ &     $   1.04  \pm   0.01  $ &     $   6.32  \pm   1.83  $ &     $   0.23  \pm   0.01  $ &     $   1.18  \pm   0.00  $ &     7173   &   87  \\ 
..&  (10.6,10.8) &$   0.86  \pm   0.02  $ &     $  10.67  \pm   0.04  $ &     $   8.96  \pm   0.02  $ &     $   0.96  \pm   0.02  $ &     $   6.56  \pm   1.95  $ &     $   0.23  \pm   0.02  $ &     $   1.22  \pm   0.00  $ &     5136   &   79  \\ 
..&  (10.8,12.0)&$   0.86  \pm   0.02  $ &     $  10.88  \pm   0.07  $ &     $   8.96  \pm   0.03  $ &     $   0.78  \pm   0.03  $ &     $   6.69  \pm   2.28  $ &     $   0.17  \pm   0.03  $ &     $   1.25  \pm   0.00  $ &     2283   &   55  \\ 
 \tableline
  \multicolumn{11}{c}{Stacks of high $g-r$ color bin ($g - r > 0.63$)}\\  
  \tableline
  (0.80,0.90)& (9.0,12.0) &$ 0.84  \pm   0.02  $ &     $  10.52  \pm   0.15  $ &     $   8.93  \pm   0.01  $ &     $   1.00  \pm   0.01  $ &     $   6.01  \pm   1.85  $ &     $   0.24  \pm   0.01  $ &     $   1.20  \pm   0.00  $ &     33646   &   194  \\ 
..& (9.0,10.0) &$   0.83  \pm   0.02  $ &     $   9.94  \pm   0.04  $ &     $   8.76  \pm   0.02  $ &     $   1.07  \pm   0.01  $ &     $   4.32  \pm   1.59  $ &     $   0.18  \pm   0.01  $ &     $   1.11  \pm   0.00  $ &     3281   &   47  \\ 
.. &(10.0,10.2)&$   0.84  \pm   0.02  $ &     $  10.12  \pm   0.04  $ &     $   8.83  \pm   0.01  $ &     $   1.05  \pm   0.01  $ &     $   4.72  \pm   1.61  $ &     $   0.19  \pm   0.01  $ &     $   1.13  \pm   0.00  $ &     7225   &   77  \\ 
..&  (10.2,10.4) &$   0.83  \pm   0.02  $ &     $  10.32  \pm   0.04  $ &     $   8.90  \pm   0.01  $ &     $   1.02  \pm   0.01  $ &     $   5.47  \pm   1.70  $ &     $   0.21  \pm   0.01  $ &     $   1.16  \pm   0.00  $ &     8548   &   90  \\ 
..& (10.4,10.6)&$   0.83  \pm   0.02  $ &     $  10.51  \pm   0.04  $ &     $   8.94  \pm   0.02  $ &     $   1.04  \pm   0.01  $ &     $   5.98  \pm   1.71  $ &     $   0.26  \pm   0.01  $ &     $   1.19  \pm   0.00  $ &     7173   &   89  \\ 
..&  (10.6,10.8) &$   0.84  \pm   0.02  $ &     $  10.69  \pm   0.04  $ &     $   8.98  \pm   0.02  $ &     $   0.93  \pm   0.02  $ &     $   6.34  \pm   1.85  $ &     $   0.24  \pm   0.02  $ &     $   1.24  \pm   0.00  $ &     5137   &   81  \\ 
..&  (10.8,12.0)&$   0.85  \pm   0.02  $ &     $  10.88  \pm   0.06  $ &     $   8.98  \pm   0.04  $ &     $   0.90  \pm   0.03  $ &     $   6.93  \pm   2.15  $ &     $   0.24  \pm   0.03  $ &     $   1.26  \pm   0.00  $ &     2283   &   59  \\ 
\enddata
\tablecomments{(1) Redshift range for stacking spectra. Except the redshift bins, the redshift is restrained to $0.8<z<0.9$ for stacking spectra in bins of other physical parameters. (2) Stellar mass range in $\log(M_\star/M_\sun)$. (3) Median redshift and corresponding standard deviation. (4) Median mass in $\log(M_\star/M_\sun)$ and corresponding standard deviation. (5) Estimated metallicity in $\rm 12 + log(O/H)$ and its error for each stacked spectrum. (6) Star formation rate in unit of logarithmic $\ M_{\odot}\rm \ Gyr^{-1}$ and its error for each stack. (7) Median half-light radius in kpc and corresponding standard deviation. (8) Calculated intrinsic dust reddening and its error in mag for each stack. (9) Calculated $D_n(4000)$ and its error for each stack. (10) Number of individual galaxies used for stacking. (11) Median S/N of all wavelength pixels in each stack. }
\end{deluxetable*}

\subsection{Effect of sample selection}
\citet{Raichoor2017} investigated the properties of the eBOSS ELG sample and presented the typical features of star-forming galaxies. Our stellar mass and SFR measurements of the composite ELG spectrum lie on the star formation sequence of star-forming galaxies at $0.5<z<1.0$ of \citet{Whitaker2014}, which also supports that the eBOSS ELGs are typical star-forming galaxies. Although we have made comparisons of our MZR and FMR with those for star-forming galaxies in other studies as consistent as possible (e.g., using identical IMF and dust extinction law), there remains the issue that sample inhomogeneity and incompleteness as well as the metallicity calibration method and measurement uncertainties might affect such comparisons. As explored by \citet{Guo2019} and \citet{Gonzalez2018}, the eBOSS ELG samples are incomplete in both stellar mass and $\oii$ luminosity functions. 

The overall MZR relation in the redshift range of $0.6<z<1.05$ presented in Figure \ref{MZR} suffers from the redshift sample inhomogeneity, which makes the most massive subsample drop down. Such inhomogeneity can be seen in the statistics of the overall redshift bin in Table \ref{tab2}, where the median redshift increases with the median mass. Thus, the MZRs in smaller redshift bins should be more accurate. \citet{Guo2019} derived the stellar mass completeness function of eBOSS ELG sample, illustrating that stellar mass completeness in $0.8<z<0.9$ is higher than those in $0.7<z<0.8$ and $0.9<z<1.0$. Therefore, Figure \ref{redshiftevolution} shows a more clear trend in $0.8<z<0.9$ than the other two redshift range. The MZR and FMR in $0.8<z<0.9$ will suffer less from the mass incompleteness. 

\begin{figure}[!tb]
\begin{center}
\includegraphics[width = 0.47\textwidth]{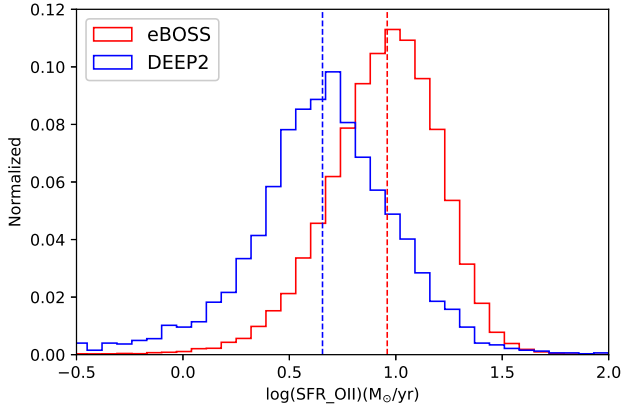}
\caption{Normalized SFR distributions for eBOSS and DEEP2 galaxy samples at $0.75 < z < 0.82$. The dashed lines denote the median values. The median difference of the distributions is about 0.3 dex.}
\label{sfrcomp}
\end{center}
\end{figure}

The sample selections are different for eBOSS ELGs and other samples at a similar redshift. From a direct comparison between Figure \ref{MZR} and \ref{redshiftevolution}, our MZR for the subsample with median $z\sim0.75$ is systematically lower than that for the sample of \citet{Zahid2014} at $z\sim0.78$, which presents an opposite trend of the MZR redshift evolution. In addition to the measurement uncertainties and different metallicity calibrations, the most important factor causing this befaviour is the sample selection difference between eBOSS and DEEP2. The eBOSS target selection favours the ELGs with relatively strong $\oii$ emission. Figure \ref{sfrcomp} gives a SFR comparison between the eBOSS and DEEP2 galaxy samples in the same redshift range of $0.75 < z < 0.82$ as used in \citet{Zahid2014}. Both SFR estimations are derived from the {\oii} luminosity measurement. The median SFR for the eBOSS ELGs is about 0.3 dex higher than that of the DEEP2 galaxies. As seen in Figure \ref{phyevo1}, a higher SFR can lead to a lower MZR.

From the target selections of DEEP2 and eBOSS, the DEEP2 samples are selected with the magnitude limit of $R<24.1$, while the eBOSS ELGs are selected with the magnitude limit of $21.8<g<22.8$. The blue color box and the bright cut ($g>21.8$) for the eBOSS ELG selection drives the incompleteness of the stellar mass at the high mass end. Moreover, as presented by \citet{Comparat2015} and \citet{Gonzalez2018}, the eBOSS ELG sample is incomplete in {\oii} luminosity functions at both bright and faint ends, produced by the $g$-band magnitude cut of $21.8<g<22.8$. The {\oii} luminosity has a strong correlation with the gas-phase metallicity. Therefore, both the incompletenesses of {\oii} luminosity and stellar mass may cause the lower plateau of our FMR relative to the local FMR.

However, the sample incompleteness would not affect our analyses of the MZR dependency on different physical parameters. Our most robust results concern the significance of the parameter dependency and the deviation of MZR between low and high parameter bins. 


\section{Summary\label{summary}}
This paper investigates the relation between stellar mass and gas-phase metallicity for galaxies at $z\sim0.8$ and explore the dependency of the MZR relation on different physical properties. The sample consists of 180,020 massive ELGs from SDSS IV/eBOSS, in the redshift range $0.6<z<1.05$ with median redshift at $0.83$, and the stellar mass covers $9 < {\rm log}(M_{\star}/M_{\odot}) < 12$. Although the S/Ns of single eBOSS spectra are low, construction of high-S/N composite spectra through stacking in different parameter bins enable us to accurately derive average properties in different parameter space and study the MZR at the specified redshift range. 

The MZR relation at $z\sim0.83$ is derived. Combining this result with other studies in the local and high-redshift universe, we confirm that the MZR has a clear redshift evolution, where more evolved galaxies have higher metallicity at a specified stellar mass. By dividing the total galaxy sample into subsamples in three redshift bins ranging from 0.6 to 1.05, we obtain the MZRs at $z\sim0.75$, 0.84, and 0.95 and also see the MZR evolution. The MZRs in smaller redshift bins should be more accurate since they suffer less from the sample inhomogeneity.  

We explore the influence of different physical parameters on the MZR including SFR/sSFR, high-light radius, mass density, and optical color. The redshift is constrained to $0.8<z<0.9$ to improve the homogeneity of the galaxy samples. The MZR moves downwards for galaxies with higher SFR/sSFR and half-light radius, while this relation increases for higher mass density and optical color. According to the amount of deviations of the MZRs obtained in different bins of each parameter, we conclude that the SFR is the most significant factor that affects MZR, and the FMR at $0.6<z<1.05$ approximately follows the local one if considering the sample inhomogeneity and incompleteness. 

The stacked spectra and the derived properties can be accessed at \url{http://batc.bao.ac.cn/~zouhu/doku.php?id=projects:mzr_ebosselg:start}.

\acknowledgments
This work is supported by the National Basic Research Program of China (973 Program; Grant Nos. 2015CB857004, 2017YFA0402600, and 2014CB845704), the National Natural Science Foundation of China (NSFC; Grant Nos. 11433005, 11673027, 11733007, 11320101002, 11421303, 11973038, and 11873032), and the External Cooperation Program of Chinese Academy of Sciences (Grant No. 114A11KYSB20160057). 

The Legacy Surveys consist of three individual and complementary projects: the Dark Energy Camera Legacy Survey (DECaLS; NOAO Proposal ID \# 2014B-0404; PIs: David Schlegel and Arjun Dey), the Beijing-Arizona Sky Survey (BASS; NOAO Proposal ID \# 2015A-0801; PIs: Zhou Xu and Xiaohui Fan), and the Mayall z-band Legacy Survey (MzLS; NOAO Proposal ID \# 2016A-0453; PI: Arjun Dey). DECaLS, BASS and MzLS together include data obtained, respectively, at the Blanco telescope, Cerro Tololo Inter-American Observatory, National Optical Astronomy Observatory (NOAO); the Bok telescope, Steward Observatory, University of Arizona; and the Mayall telescope, Kitt Peak National Observatory, NOAO. The Legacy Surveys project is honored to be permitted to conduct astronomical research on Iolkam Du'ag (Kitt Peak), a mountain with particular significance to the Tohono O'odham Nation. The Legacy Survey team makes use of data products from the Near-Earth Object Wide-field Infrared Survey Explorer (NEOWISE), which is a project of the Jet Propulsion Laboratory/California Institute of Technology. NEOWISE is funded by the National Aeronautics and Space Administration.

Funding for the Sloan Digital Sky Survey IV has been provided by the Alfred P. Sloan Foundation, the U.S. Department of Energy Office of Science, and the Participating Institutions. SDSS acknowledges support and resources from the Center for High-Performance Computing at the University of Utah. The SDSS web site is www.sdss.org.

SDSS is managed by the Astrophysical Research Consortium for the Participating Institutions of the SDSS Collaboration including the Brazilian Participation Group, the Carnegie Institution for Science, Carnegie Mellon University, the Chilean Participation Group, the French Participation Group, Harvard-Smithsonian Center for Astrophysics, Instituto de Astrofísica de Canarias, The Johns Hopkins University, Kavli Institute for the Physics and Mathematics of the Universe (IPMU) / University of Tokyo, the Korean Participation Group, Lawrence Berkeley National Laboratory, Leibniz Institut für Astrophysik Potsdam (AIP), Max-Planck-Institut für Astronomie (MPIA Heidelberg), Max-Planck-Institut für Astrophysik (MPA Garching), Max-Planck-Institut für Extraterrestrische Physik (MPE), National Astronomical Observatories of China, New Mexico State University, New York University, University of Notre Dame, Observatório Nacional / MCTI, The Ohio State University, Pennsylvania State University, Shanghai Astronomical Observatory, United Kingdom Participation Group, Universidad Nacional Autónoma de México, University of Arizona, University of Colorado Boulder, University of Oxford, University of Portsmouth, University of Utah, University of Virginia, University of Washington, University of Wisconsin, Vanderbilt University, and Yale University.

\end{document}